\documentclass[twocolumn,aps,prc,showpacs,superscriptaddress,preprintnumbers,floatfix,nofootinbib]{revtex4}
\usepackage{epsfig,graphics}
\usepackage{graphicx}
\usepackage{dcolumn}
\usepackage{bm}
\usepackage{amsmath}
\usepackage{amssymb}
\usepackage[usenames]{color}
\usepackage{ulem} 
\usepackage[colorlinks,linkcolor=blue,urlcolor=blue,citecolor=blue]{hyperref}

\usepackage{makeidx}
\makeindex
\usepackage{CJK}

\begin{document}

\begin{CJK*}{GB}{gbsn}

\title{Direct photon emission and influence of dynamical wave packets in an extended quantum molecular dynamics model}

\author{C. Z. Shi (施晨钟)}
\affiliation{Shanghai Institute of Applied Physics, Chinese Academy of Sciences, Shanghai 201800, China}
\affiliation{Key Laboratory of Nuclear Physics and Ion-beam Application (MOE), Institute of Modern Physics, Fudan University, Shanghai 200433, China}
\affiliation{University of the Chinese Academy of Sciences, Beijing 100080, China}

\author{Y. G. Ma (马余刚)}\thanks{Author to whom all correspondence should be addressed. Email: mayugang@fudan.edu.cn}
\affiliation{Key Laboratory of Nuclear Physics and Ion-beam Application (MOE), Institute of Modern Physics, Fudan University, Shanghai 200433, China}
\affiliation{Shanghai Institute of Applied Physics, Chinese Academy of Sciences, Shanghai 201800, China}

\author{X. G. Cao (曹喜光)}
\affiliation{Shanghai Advanced Research Institute, Chinese Academy of Sciences, Shanghai 201210, China}
\affiliation{Shanghai Institute of Applied Physics, Chinese Academy of Sciences, Shanghai 201800, China}

\author{D. Q. Fang (方德清)}
\affiliation{Key Laboratory of Nuclear Physics and Ion-beam Application (MOE), Institute of Modern Physics, Fudan University, Shanghai 200433, China}
\affiliation{Shanghai Institute of Applied Physics, Chinese Academy of Sciences, Shanghai 201800, China}

\author{W. B. He (何万兵)}
\affiliation{Key Laboratory of Nuclear Physics and Ion-beam Application (MOE), Institute of Modern Physics, Fudan University, Shanghai 200433, China}

\author{C. Zhong (钟晨 )}
\affiliation{Key Laboratory of Nuclear Physics and Ion-beam Application (MOE), Institute of Modern Physics, Fudan University, Shanghai 200433, China}

\date{ \today}

\begin{abstract}
Direct photon produced from first proton-neutron ($p$-$n$) collision during the early stage of heavy ion reaction is a sensitive probe to reflect energy and momentum distribution of nucleons. In this work, we embedded the hard photon production channel in an extended quantum molecular dynamics (EQMD) model, and  took the direct photon as a possible probe to improve namely the Fermi motion in the  EQMD model. A possible scheme is offered to handle the dynamical wave packet width within incoherent bremsstrahlung process. Direct photons calculated by our modified EQMD were compared with data of $^{14}$N + $^{12}$C at beam energies $E/A$ = 20, 30 and 40 MeV,
and it is found that the yield, inverse slope and angular distribution of direct photons could be reasonably reproduced. In addition, asymmetric reaction systems of $^{4}$He + C and $^{4}$He + Zn at $E/A$ = 53 MeV
are also simulated in this work. It is found that the symmetric angular distribution in
the nucleon-nucleon ($N$-$N$) center-of-mass (c.m.) frame and the velocity of $ the \gamma$-emission source can be reasonably obtained from our method although there is some  quantitative differences.
\end{abstract}

\pacs{21.65.Ef, 25.70.Mn, 21.65.Cd}

\maketitle

\section{Introduction}
Photon as an untwisted probe offers an attractive alternative way to investigate hadronic  property via intermediate energy heavy ion collisions \cite{Bonasera,CassingProduction,SchutzHard,NifeneckerHigh,N_C,He_C_Zn,DengXG1,DengXG2}. Based on this point, there was a lot of researches on photons in the past several decades. Around the Coulomb barrier about 0.1 - 10 MeV, hot nucleus is created and  its de-excited $\gamma$-ray is produced.  Above that energy, nuclear collective modes emerge. For instance,  an additional continuum $\gamma$-radiation component, namely giant dipole resonance (GDR), contributes to $\gamma$ spectra around $E_\gamma \simeq$ 10 - 25 MeV. However,  $\gamma$-ray spectra do not cutoff above the GDR region, they show a continuum high energy radiation approximately extending to the 100 MeV level for  heavy ion collisions in the regime of Fermi energy \cite{Kafexhiu}. Usually, the high energy component of $\gamma$-ray spectra is called hard photons, which are the main research subject in this article. It was suggested in early years, for instance, incoherent nucleon-nucleon bremsstrahlung \cite{incoherent, PhysRevC.34.2127}, coherent nucleus-nucleus bremsstrahlung \cite{Vasak_1985, VASAK1986276}, and nucleon bremsstrahlung in the mean-field potential \cite{BAUER1986159}, etc., are main production mechanisms of hard photons. It was experimentally demonstrated that a symmetric angular distribution of the hard photons in the $N$-$N$ c.m. frame and the velocity of the $\gamma$-emission source is close to half of the beam's velocity. Such a fact strongly suggests the incoherent proton-neutron ($p$-$n$) collision during  the early stage of nuclear interaction is dominated for the hard photon bremsstrahlung \cite{BERTHOLET1987541, BIRO1987579,LiuGH2008,MaYG2012,YongGC}. Based on  this mechanism, the inverse slope of hard photons could be taken as  a measure of the energy and momentum distributions of nucleons which are responsible for hard photon production \cite{SchutzHard, CassingProduction,Tam_incoherent}.

On the other hand, the $\alpha$-clustering phenomenon as a novel nuclear structure has received great attention in recent years. An extended quantum molecular dynamics model \cite{EQMD}, as one of a few microscopic transport models which can give $\alpha$-clusters with a nice computation performance,  has  succeeded in describing multifragmentation \cite{hagel}, giant dipole resonance \cite{He_wb_gdr_prl,HeWB2,WangSS,WangK}, photonuclear reactions \cite{Huang_bs} as well as collective flow and shear viscosity etc \cite{guo_cc,GuoCQ} at Fermi energy. In comparison with the traditional QMD-type model and its many applications \cite{Aichelin,Hartnack1998Modelling,Yan1,WangTT,Yan2,LiPC,Ono,ZhangZF,FengZQ,Sood,Desai,LiuHL},
the EQMD model has been improved in some aspects. For example, a phenomenological Pauli potential was added, the dynamical degree of wave packets was considered, and a friction cooling method for the initialization of nuclei was used. Although EQMD has some advantages to describe novel structures like $\alpha$-clusters, the effect of dynamic wave packets, such as Fermi motion in the original article  \cite{EQMD}, has not yet fully taken into account the model. Even though it has no significant influence on elastic scattering between nucleons, it is very important to the inelastic process, e.g., inherent proton-neutron bremsstrahlung.

In this work, a new method is suggested to deal with inelastic scattering including the dynamical wave packet effect specifically for such an EQMD model. We shall show some results calculated by our modified EQMD model for the hard photon production around the Fermi energy.

The rest of the paper is organized as follows. The new method is briefly introduced in Sec. \ref{sec:re}. The detailed comparisons between the simulation results and experimental data are given in Sec. \ref{sec:result}.  A summary is given in Sec. \ref{sec:con}.

\section{Re-extract nucleon's kinetic energy considering dynamic wave packet effect}
\label{sec:re}

\subsection{Direct photon}
\label{direct_photon}

Because of a destructive interference effect, bremsstrahlung in proton-proton collisions is an order of magnitude lower than the proton-neutron collision \cite{METAG1988483, DENTERRIA200227}. In this paper, only the first instant proton-neutron collision, i.e. $pn \rightarrow pn\gamma$, is taken into account in the EQMD model. This elementary cross section adopts the hard-sphere limit and it is consistent with energy conservation from Bauer {\it et al.} \cite{PhysRevC.34.2127},
\begin{equation}
\frac{d^{2} \sigma^{\text {elem }}}{d E_{\gamma} d \Omega_{\gamma}} = \alpha_c \frac{R^{2}}{12 \pi} \frac{1}{E_{\gamma}}\left(2 \mathbf{\beta}_{f}^{2}+3 \sin ^{2} \theta_{\gamma} \mathbf{\beta}_{i}^{2}\right).
\label{eq_element}
\end{equation}
Here $R$ is the radius of the hard sphere, $\alpha_c$ is the fine structure constant, $\mathbf{\beta}_{i}$ and $\mathbf{\beta}_{f}$ are  the initial and final velocity of the proton in the proton-neutron center-of-mass system, and $\theta_\gamma$ is the angle between the incident proton direction and the photon emitting direction. More details of the model can be found in Refs. ~\cite{PhysRevC.34.2127,LiuGH2008}.

The probability for emitting a photon in a single $p$-$n$ collision can be written as
\begin{equation}
\begin{aligned}
\int & \frac{d \Omega_{e}}{4 \pi} \frac{1}{\sigma_{N N}} \frac{d^{2} \sigma^{\text {elem }}}{d E_{\gamma} d \Omega_{\gamma}} \\
& \times\left[1-S_{3}\left(\mathbf{r}_3, \mathbf{k}_{3}, t\right)\right]\left[1-S_{4}\left(\mathbf{r}_4, \mathbf{k}_{4}, t\right)\right].
\end{aligned}
\label{eq_probibility}
\end{equation}
Here $\sigma_{NN}$ is the elemental nucleon-nucleon cross section, $\mathbf{r}_3$, $\mathbf{k}_3$ and $\mathbf{r}_4$, $\mathbf{k}_4$ are the coordinates and momenta of the scattered protons and neutrons. The quantities $S_{3}$ and $S_{4}$ are the effective occupation fraction of phase space of the scattered particles, and the term $[1-S_{3}(\mathbf{r}_3,\mathbf{k}_3,t)][1-S_{4}(\mathbf{r}_4,\mathbf{k}_4,t)]$ represents the effect of Pauli blocking in the final state. $\Omega_e$ is the solid angle of the vector $\mathbf{k}_3-\mathbf{k}_4$. Because $\Omega_e$ can not be uniquely determined by the conservation of energy and momentum, it is necessary to integrate $4\pi$ solid angle to obtain the occupation in the final state.

Nowadays, it is generally believed that the hard photon cross section can be parametrized as
\begin{equation}\label{photon_section}
  \sigma_\gamma =\sigma_{R} N_{np} P_{\gamma},
\end{equation}
where $\sigma_{R}$ is the reaction cross section, $N_{np}$ is the total number of initial $np$ collisions averaged over the impact parameter, and $P_\gamma = C \times exp\left( -\frac{E^{min}_{\gamma}}{E_{0}}\right)$ is the hard photon ($E_{\gamma}>$30MeV) emission probability in a single collision. $\sigma_R$ can be easily evaluated by the maximum impact parameter. The $N_{np}$ can be calculated by geometrical equal-participant model and its value depends on the proton and neutron numbers of projectile and target nuclei. $C$ is a constant  fitted to experimental data, $E^{min}_{\gamma}=30$MeV and $E_0$ is the inverse slope which depends on the N-N collision energy. The inverse slope can be obtained by fitting the hard photon energy spectrum whose sharpness is close to exponential distribution $\frac{d\sigma_{\gamma}}{dE_{\gamma}}=K exp\left( - \frac{E_{\gamma}}{E_0} \right)$.
Although there are many factors that affect the yield, the inverse slope is an essential physical quantity related to collision energy.  Based on this reason, we use hard photon to reflect the momentum distribution in the nucleus when the incident energy is known.

After the first compression stage during heavy ion collisions, an additional thermal photon emission source could appear \cite{thermal_photon_1, MARTINEZ199523,ZhangF}.  This is from the secondary $p$-$n$ collisions within thermalizing zone. However, this component is not important for light collision systems \cite{Kafexhiu} because of  not enough $p$-$n$ collisions \cite{thermal_photon_1}. Based on this reason, only the direct photons are considered in this paper to avoid discussing the lifetime of the hot zone in the EQMD model. In some degrees, this truncation will probably underestimate the yield of hard photons at the low energy edge which are not important in this work.

\subsection{EQMD model}
\label{eqmd_modle}

EQMD model is a many-body transport   model in which  each nucleon is represented by a Gaussian wave packet, and the total wave function as a simple direct product of those wave packets, can be written as \cite{EQMD}
\begin{equation}
\begin{aligned}
\Psi=&\prod_i\phi_i(\mathbf{r}_i)\\
=&\prod_i{\left( {\frac{{{v_i} + v_i^ * }}{{2\pi }}} \right)^{3/4}}\exp \left[ { - \frac{{{v_i}}}{2}{{\left( {{{\bf r}_i} - {{\bf R}_i}} \right)}^2} + \frac{i}{\hbar }{{\bf P}_i} \cdot {{\bf r}_i}} \right],\\
\end{aligned}
\label{eq_eqmd}
\end{equation}
where $\mathbf{R}_i$ and $\mathbf{P}_i$ are the centers of position and momentum of the $i$-th nucleon (wave packet), $\upsilon_i = \frac{1}{\lambda_i} + i\delta_i$ is the complex width of the dynamic wave packets, $\lambda_{i}$ and $\delta_{i}$ are dynamic variables and represent the real and imaginary parts, respectively.
The ${v_i}$  for each nucleon is dynamic and independent.

The equations of motion with 4$A$ parameters ($A$ means the number of constituent particles) are following the time-dependent variation principle (TDVP) \cite{FMD,KERMAN1976332}, and they can be written as follows \cite{EQMD},
\begin{equation}
\begin{aligned}
\dot{\mathbf{R}}_{i} = \frac{\partial H}{\partial \mathbf{P}_{i}}+\mu_{\mathrm{R}} \frac{\partial H}{\partial \mathbf{R}_{i}}, ~~~& \dot{\mathbf{P}}_{i} = -\frac{\partial H}{\partial \mathbf{R}_{i}}+\mu_{\mathrm{P}} \frac{\partial H}{\partial \mathbf{P}_{i}} \\
\frac{3 \hbar}{4} \dot{\lambda}_{i} = -\frac{\partial H}{\partial \delta_{i}}+\mu_{\lambda} \frac{\partial H}{\partial \lambda_{i}}, ~~~& \frac{3 \hbar}{4} \dot{\delta}_{i} = \frac{\partial H}{\partial \lambda_{i}}+\mu_{\delta} \frac{\partial H}{\partial \delta_{i}}.
\end{aligned}
\label{eq_motion}
\end{equation}
Here $\mu_{\mathbf{R}}$, $\mu_{\mathbf{P}}$, $\mu_{\lambda}$ and $\mu_{\delta}$ are friction coefficients, and $H$ is the Hamiltonian.
The system dissipates its energy and evolves to a stable (minimum or even eigenstate) state with negative values of these friction coefficients in initialization after taking several thousand $fm/c$. As a result, a  very stable nucleon distribution can be obtained. Contrarily, the total energy can keep stable with zero value of these friction coefficients during the nucleus-nucleus  ($n$-$n$) collision. It must be noted that there is a certain arbitrariness in the selection of the friction coefficients and cooling time to some extent, however,  we will not discuss those skills here.

Usually, the width of the wave packet depending on system size is a constant during the reaction in most QMD-type models, its contribution on kinetic energy was  subtracted. However, in the EQMD model, it takes into account the wave packet width degree of freedom, and there is no classical physical quantity corresponding to those wave packets.

The Hamiltonian $H$ is written as
\begin{equation}
\begin{aligned}
H & = \left\langle\Psi\left|\sum_{i}-\frac{\hbar^{2}}{2 m} \nabla_{i}^{2}-\hat{T}_{\mathrm{zero}}+\hat{H}_{\mathrm{int}}\right| \Psi\right\rangle \\
& = \sum_{i}\left[\frac{\mathbf{P}_{i}^{2}}{2 m}+\frac{3 \hbar^{2}\left(1+\lambda_{i}^{2} \delta_{i}^{2}\right)}{4 m \lambda_{i}}\right]-T_{\mathrm{zero}}+H_{\mathrm{int}}.
\end{aligned}
\label{eq_h}
\end{equation}
The second and third terms represent the kinetic energy from the momentum variance of wave packets and  the spurious zero-point center-of-mass kinetic  energy $T_{zero}$  \cite{zero_energy}, $H_{int}$ represents the effective interaction potential energy  which will be described later.
The kinetic energy including two parts is a pure quantum effect,
which can be easily understood as follow
\begin{equation}
\frac{\hat{\langle \mathbf{p}_{i}^{2}\rangle}}{2m_i}-\frac{\langle \hat{\mathbf{p}_{i}}\rangle^{2}}{2m_i}\neq0.
\end{equation}
Here $\frac{\langle\hat{\mathbf{p}_{i}^{2}}\rangle}{2m_i}$ and $\frac{\langle\hat{\mathbf{p}_i}\rangle^{2}}{2m_i}$ are the $i$-th nucleon's kinetic energy and the center of $i$-th wave packets' kinetic energy. Their difference is not equal to zero because  a nucleon is treated as a wave packet rather than a point particle.

For two-body collision part, only elastic scattering was considered in EQMD model. Same as most QMD-type models, EQMD uses the center of wave packets from the phase space to represent particles within scattering process. If a collision is accepted, their momenta are changed, but the position and the width of wave packet are remained. In addition, the system should satisfy the energy-conservation and fermionic properties at the final state. If a Pauli-blocking occurs, it  restores them to the previous state.

Generally speaking, it is reasonable for most transport models to treat the nucleon as a point particle, same as the cascade model, rather than a wave packet during two-body collision processing. It is also reasonable for the EQMD model to treat elastic scattering in most cases, however, the disadvantage will emerge for treating inelastic scatterings because of the inconsistency of the particles propagation and two-body collision. The performance of this disadvantage will be shown in the next section.

For the effective interaction $H_{int}$ which was mentioned in Eq.~\ref{eq_h}, it consists of the Skyrme and  Coulomb potentials, the symmetry energy, and the Pauli potential as follows:
\begin{equation}
H_{int} = H_{Sky.} + H_{Coul.} + H_{Sym.} + H_{Pauli}.
\end{equation}
The form of Skyrme interaction used in the EQMD model is the simplest, written as
\begin{eqnarray}
H_{Sky.} = \frac{\alpha }{2\rho _{0}}\int \rho^2\left ( \mathbf{r} \right )d^3 r+\frac{\beta }{\left ( \gamma +1 \right )\rho _{0 }^{\gamma}}\int \rho^{\gamma +1}\left ( \mathbf{r} \right )d^3 r,
\end{eqnarray}
where $\alpha$, $\beta$, and $\gamma$  are potential parameters which are listed in Table. I.
The symmetry potential is written as
\begin{eqnarray}
H_{Sym.} = \frac{C_{S}}{2\rho _{0}}\sum_{i,j\neq i} \int \left [ 2\delta \left ( T_i,T_j \right )-1 \right ]\rho_i\left ( \mathbf{r} \right )\rho_j\left ( \mathbf{r} \right )d^3r,
\end{eqnarray}
where $C_S$ is the symmetry energy coefficient.
Specifically, the Pauli potential is presented as
\begin{eqnarray}
H_{Pauli} = \frac{c_P}{2}\sum_i(f_i-f_0)^\mu\theta(f_i-f_0),\\
f_i\equiv\sum_j\delta(S_i,S_j)\delta(T_i,T_j)|\langle\phi_i|\phi_j\rangle|^2,
\end{eqnarray}
where, $f_i$ is the overlap of a nucleon $i$ with nucleons having the same spin and isospin, $\theta$ is the unit step function, $c_P$ is a coefficient related to strength of the Pauli potential, $f_0$ and $\mu$ are parameters.
This potential inhibits the system from collapsing into the Pauli-blocked state at low energy and gives the model capability to describe $\alpha$-clustering. Table 1 gives two parameter sets in the EQMD model \cite{EQMD}, and we take the set 2 in this work.

\begin{center}
\scriptsize
Table 1. Parameter values used in the Skyrme, symmetry, and Pauli potentials\\
\begin{tabular}{|c|c|c|c|c|c|c|c|}\hline
Parameter & $\alpha$ & $\beta$&$\gamma$&$c_{S}$&$c_{P}$&$f_{0}$ & $\mu$ \\
sets & (MeV) & (MeV) & & (MeV) & (MeV) & & \\\hline
Parameter set 1 &-116.6 &70.8 &2 &25 &15 &1.05 &2.0\\\hline
Parameter set 2 &-124.3 &70.5 &2 &25 &15 &1.0  &1.3\\\hline
\end{tabular}\\
\end{center}

\subsection{Re-extract kinetic energy within inelastic scattering}
\label{re-extract_energy}

The Wigner function of the EQMD model is written as
\begin{equation}
\begin{aligned}
w(\mathbf{r}, \mathbf{p})& = \left(\frac{1}{\pi \hbar}\right)^{3} \sum_{i}^{A} \exp [ -\frac{1+\lambda_{i}^{2} \delta_{i}^{2}}{\lambda_{i}}\left(\mathbf{r}-\mathbf{R}_{i}\right )^{2}\\
&-\frac{2 \lambda_{i} \delta_{i}}{\hbar}\left(\mathbf{r}-\mathbf{R}_{i}\right)\left(\mathbf{p}-
\mathbf{P}_{i}\right) -\frac{\lambda_{i}}{\hbar^{2}}\left(\mathbf{p}-\mathbf{P}_{i}\right)^{2}],  
\end{aligned}
\label{eq_wigner}
\end{equation}
this quantity $w(\mathbf{r},\mathbf{p})$ is the closest analog to classical phase-space density \cite{BUU}, and it does not produce negative value in our case.
We assume that the $i$-th nucleon's coordinate position is still $\mathbf{R}_i$ during two-body scattering,
then it is easy to obtain the conditional probability for the $i$-th nucleon momentum distribution
\begin{equation}
\begin{aligned}
w_{i}\left(\mathbf{p}\right) = &w_{i}\left(\mathbf{R}_{i}, \mathbf{p}\right) / \rho_{i}\left(\mathbf{R}_{i}\right)\\
= &\left(\frac{\lambda_{i}}{\pi \hbar^{2}}\right)^{3 / 2} \exp \left\{-\frac{\lambda_{i}}{\hbar^{2}}\left(\mathbf{p}-\mathbf{P}_{i}\right)^{2}\right\}.
\end{aligned}
\label{momentum_distribution}
\end{equation}
Here $w_i(\mathbf{p})$ represents the probability of finding the $i$-th nucleon's momentum with $\mathbf{p}$ when its position is  known at $\mathbf{R}_i$. Now we sample the $i$-th nucleon's momentum randomly in a single collision,
\begin{equation}
\begin{aligned}
&\mathbf{p}_{i} = \mathbf{P}_{i}+\Delta \mathbf{p} \times \sqrt{1-\frac{1}{M_i}}.
\end{aligned}
\label{momentum_sample}
\end{equation}
Here $\Delta \mathbf{p}$ is a random value given along its momentum distribution. The term including $M_{i}$, namely ``mass number" of the fragment to which the wave packet $i$ belongs \cite{EQMD}, is taking account of zero-point center-of-mass kinetic energy. The detailed definition can be found in Ref.~\cite{EQMD}.

It has to be noted that $\Delta \mathbf{p}$ is sampled only once within a single collision for simplified calculation. Strictly speaking, it is necessary to sample adequately to obtain a smooth momentum distribution. However, this deviation will be suppressed as the increase of event numbers.
Another potential risk is the unrestricted sampling. It may introduce a huge collision energy which is nonphysical. In the view of the above points, a method of adequately sampling with an appropriate cutoff which is similar to the test-particle method used in BUU-type model maybe a better choose. Fortunately, the results from our simulations shown in the next section are enough to provide some useful information.

Specifically, the wave packets will vanish after sampling since we use points instead of packets.
It is necessary to modify the Pauli-blocking treatment from the original EQMD.
In most QMD-type models, the occupation number of the $i$-th nucleon can be calculated by integrating the Wigner function on a hypercube of volume $h^3$ in the phase space centered around the point $\left( \mathbf{R}_i,\mathbf{P}_i\right)$.
If the occupation is bigger than $1$, Pauli-blocking happens. Otherwise, it performs a reject sampling to decide whether this collision can occur or not. In this work, we take directly the product of volume $h^3$ and the phase space density of the points where the scattering particle is in the final state to replace the integral of the Wigner function because of the loss of wave packet information after sampling. The new effective occupation of the $i$-th nucleon in the final state can be defined as follows
\begin{equation}
S_i = h^3 \times\sum_{j \ne i}\delta_{\tau_{i},\tau_{j}} \delta_{s_i,s_j} w_j(\mathbf{R}'_i,\mathbf{P}'_i).
\end{equation}
Here $w_j$ is the density of the Wigner function contributed by spectator nucleons  to the point $(\mathbf{R}'_i,\mathbf{P}'_i)$ at where the scattered nucleon is the final state. Compared with the conventional version, this new definition is nonideal in some degrees. It might cause more fluctuation on the photon yield. However, it is a feasible and simple way to include this fermonic property based on the actual conditions described above.

In this section, we have introduced a possible scheme to treat inelastic collision which is consistent with the momentum variance of wave packets in EQMD. However, the point particles would diffuse after a scattering, in other word, their wave packet can not be recovered. The treatment of inelastic scattering described above can only deal with the relatively rare process which can be treated as a perturbative process, i.e. hard photon bremsstrahlung.
Here we should emphasize that, this new method does not add extra energy into the model, it only re-extracts collision energy which is in consistent with the wave packet effect of the particle itself.

\section{Results and Discussion}
\label{sec:result}

\begin{figure}[b]
\resizebox{8.6cm}{!}{\includegraphics{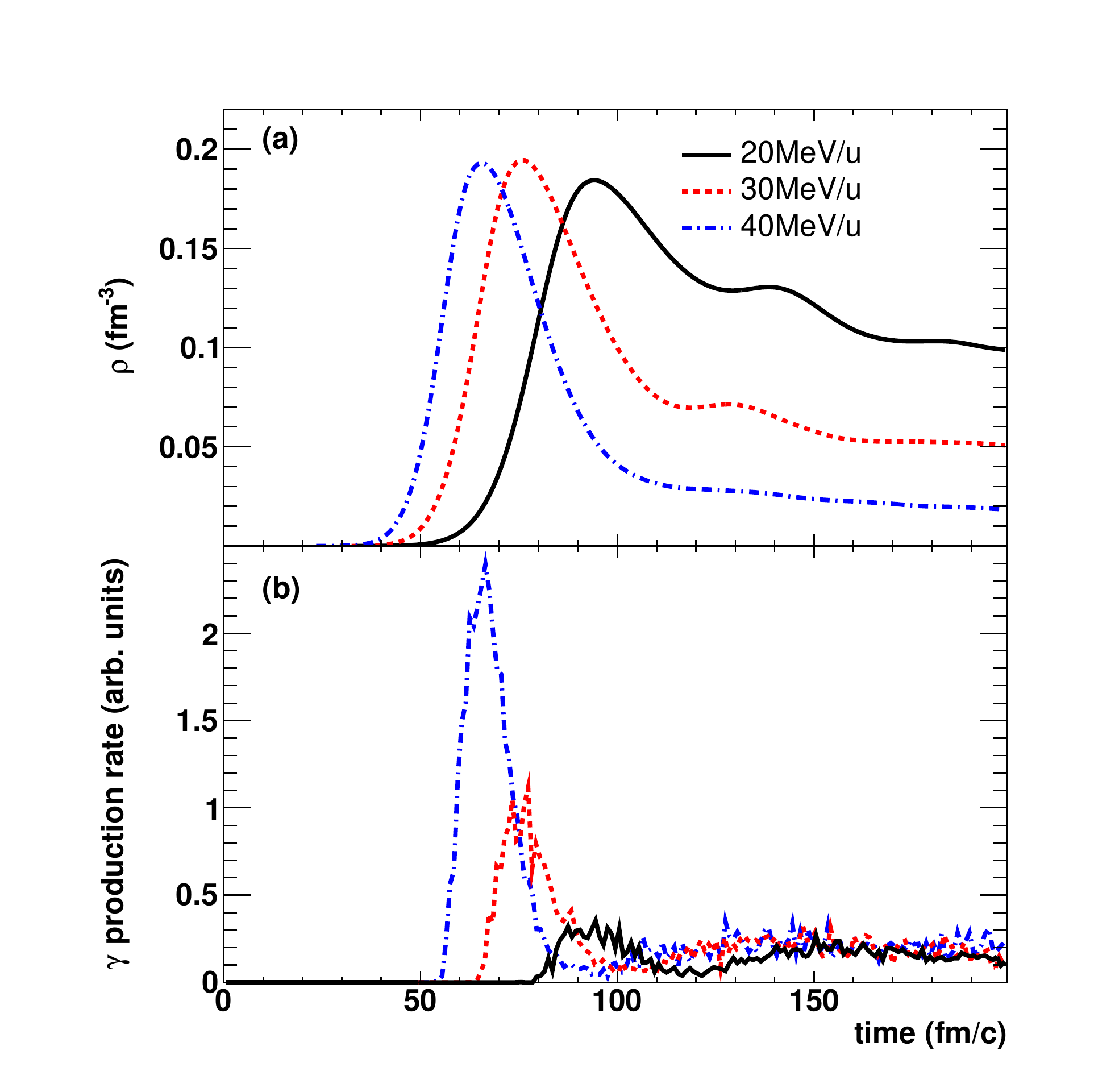}}
 \caption{Time evolution of density at point (0,0,0)  in $n$-$n$ c.m. frame (a) and hard photon production rate  (b) for $^{14}$N + $^{12}$C collisions at incident energy $E/A$ = 20 (black solid line), 30 (red dotted line) and 40 (blue dash dot line) MeV.
 }
\label{yield_time}
\end{figure}

Figure \ref{yield_time} shows time evolution of  density at point (0,0,0) in the $n$-$n$ c.m. frame (a) as well as hard photon ($E_\gamma>$ 30 MeV) production rate (b) for $^{14}$N + $^{12}$C at incident energy $E/A$ = 20 (black solid line), 30 (red dotted line) and 40 (blue dash-dot line) MeV. The photon yield was calculated by the modified EQMD model. Here we  emphasize the $np \rightarrow np\gamma$ channel is embedded in the EQMD model as a perturbative process by this work.
In the EQMD simulation, we choose 20fm as the initial distance between projectile and target nuclei.
Figure ~\ref{yield_time} displays a strong synchronization with time evolution of system compression. Shortly after  approaching  two nuclei, i.e. almost at the staring point of each curve in Fig. \ref{yield_time} (a), the production of the direct photon rapidly increases, and reaches the maximum value when the participated zone has the maximum overlap. During the later stage, the production drops fast until  the separation of target-like and projectile-like fragments. Meanwhile in the central zone, a quasiequilibrium  thermal  zone could be formed and thermal photons shall be created by a similar mechanism, but the magnitude and inverse slope are much smaller than the direct photon component \cite{MARTINEZ199523}. As mentioned before,
the emission of thermal photons and the living time of the hot zone which are supposed to be correlated with incompressibility (EOS) \cite{MARTINEZ199523,MaYG2012} are not the goal of this work, therefore we only select the early stage direct photons to discuss in this paper. The time interval for direct photon production of 75-110, 65-105, 55-95 fm/c are selected for incident energy $E/A$ = 20, 30 and 40 MeV, respectively.

\begin{figure}[htbp]
\resizebox{8.6cm}{!}{\includegraphics{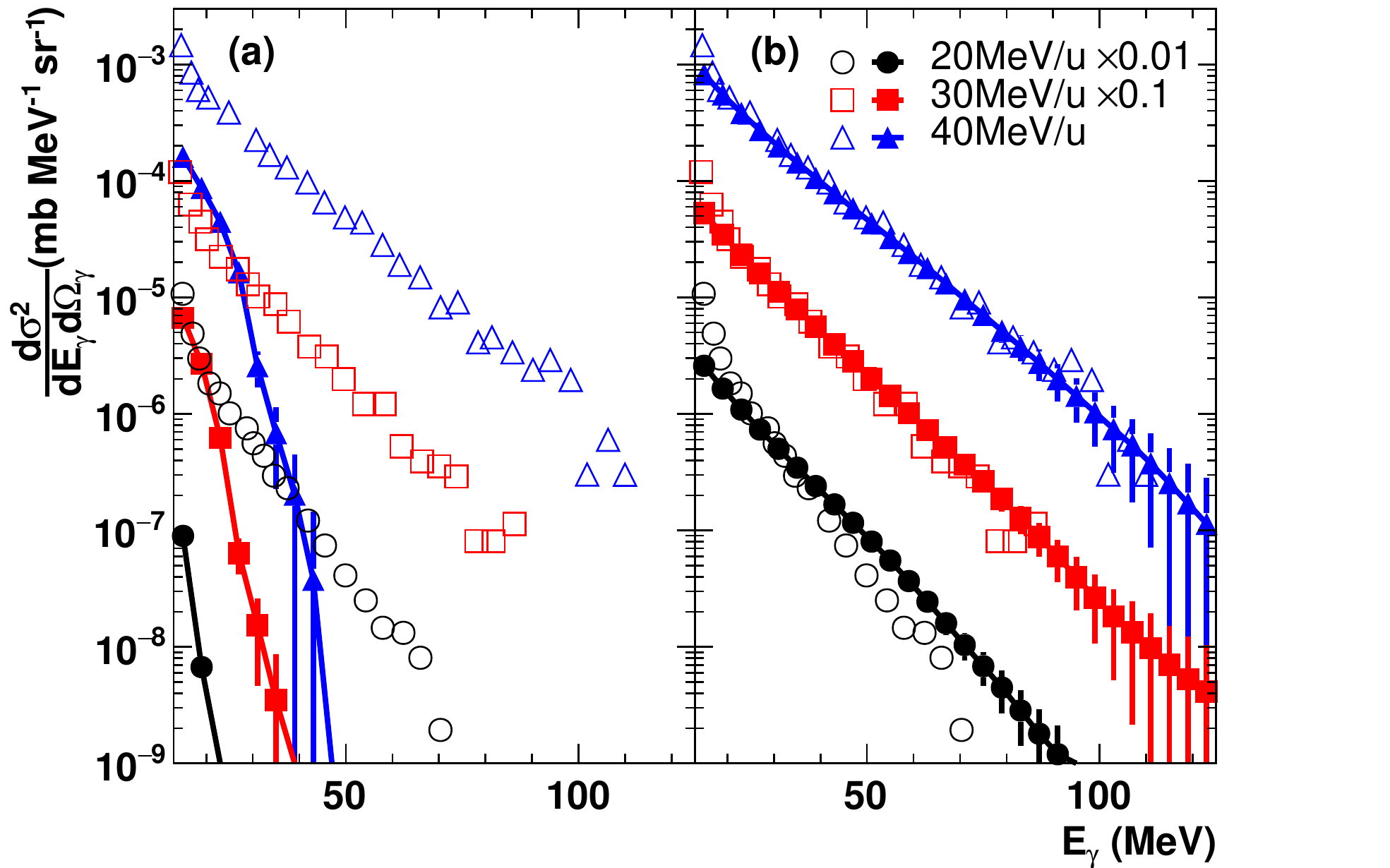}}
\caption{Photon energy spectra from $^{14}$N + $^{12}$C at incident energy $E/A$ = 20 (black), 30 (red) and 40 (blue) MeV. The solid lines with symbols in  (a) are the results calculated by the original EQMD model, while the solid lines with symbols  in  (b) are the results simulated by the modified EQMD model. The open markers represent the experimental data taken from Ref.~\cite{N_C}. The selected polar angle of those photons is about $\theta_{lab} \approx 90^{\circ}$ ($-0.1<cos(\theta_{lab})<0.1$) with respect to the beam axis.
}
\label{energy_n_c}
\end{figure}

\begin{figure}[htbp]
\resizebox{8.6cm}{!}{\includegraphics{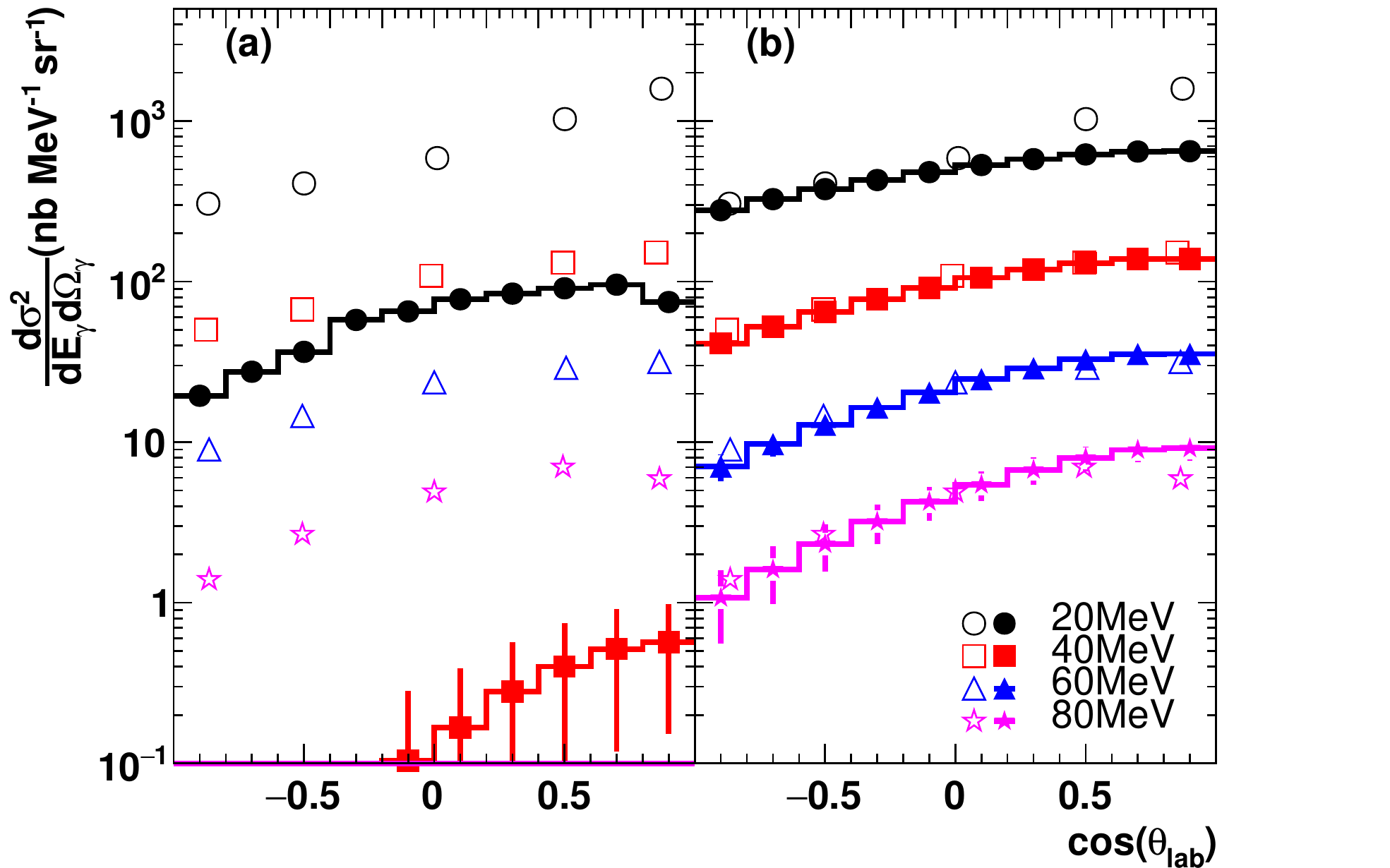}}
\caption{
Laboratory angular distributions of photons for  40 MeV/nucleon $^{14}$N + $^{12}$C, at photon energies of $E_\gamma$ = 20 $\pm$ 3 (dark), 40 $\pm$ 3 (red), 60 $\pm$ 3 (blue) and 80 $\pm$ 3 (pink)  MeV. The  lines with solid symbols in panel (a) are the results calculated by the original EQMD and the   lines with solid symbols in panel (b) are the results simulated by the modified EQMD model. The  open markers represent the experimental data taken from Ref.~\cite{N_C}.}
\label{angle_n_c}
\end{figure}

Figure \ref{energy_n_c} shows energy spectra of hard photons within about polar angle $\theta_{lab} \approx 90^{\circ}$ for $^{14}$N + $^{12}$C at incident energy $E/A$ = 20 (black), 30 (red) and 40 (blue) MeV. The results calculated by the original EQMD are plotted in panel (a) and the results from the modified EQMD are shown  in panel (b).
The experimental data (open markers) from Steven {\it et al.} \cite{N_C} show an exponential fall off with a constant inverse slope in the hard photon energy spectra above $E_\gamma >$ 30 MeV.  It is clearly seen that there is a significant deviation on magnitude as well as the inverse slope of  hard photon energy spectra calculated by the original EQMD model.
The limit of the highest energy of direct photons by the original EQMD is about 21, 36 and 45 MeV at incident energy $E/A$ = 20, 30 and 40 MeV, respectively. Because of the  first $p$-$n$ collision is dominant in incoherent bremsstrahlung process, the upper limit of hard photon energy directly reflects the maximum available collision energy. The tail of the high energy photons from experimental data indicates that the energies of nucleon have been underestimated   in the original model calculations. Those results show the defect of the conventional two-body collision method to deal with  inelastic process in the EQMD model. In contrast, from our modified EQMD model, both the magnitude and inverse slope of the experimental data can be reasonably reproduced. At  low energy parts with $E_\gamma<$ 30 MeV, the deviation increases as  photon energy decreases because  other mechanisms could rise up in this range. At the high energy parts, the error certainly increases because of fewer number of high energy scattering nucleons. It is easy to understand from  Eq.~\ref{momentum_distribution}, the probability of finding a nucleon in momentum space is rapidly dissipated if it is far away from the center of the Gaussian wave packet.

Figure 3 shows the laboratory angular distributions of photons for 40 MeV/nucleon $^{14}$N + $^{12}$C with photon energies of $E_{\gamma} = 20 \pm 3$  (dark), $40  \pm 3$ (red), $60  \pm 3 $ (blue) and $80  \pm 3 $ (pink) MeV. The results calculated by the original EQMD (solid line) are plotted in Fig. 3(a) and the results simulated by the modified EQMD (solid line) are plotted in Fig. 3(b). The angular distributions from the same experiment (open markers) are all slightly forward peaked. It is from the $\gamma$-emission source having half of the beam velocity. Because of the limitation of the original EQMD model in calculating the hard photon, the spectra of angular distribution above energy $E_\gamma \geq 40$MeV is much smaller than the experimental value. On the contrary, both the magnitude and the shape of angular distribution are reasonably reproduced using the modified EQMD model.

\begin{figure}[b]
\resizebox{8.6cm}{!}{\includegraphics{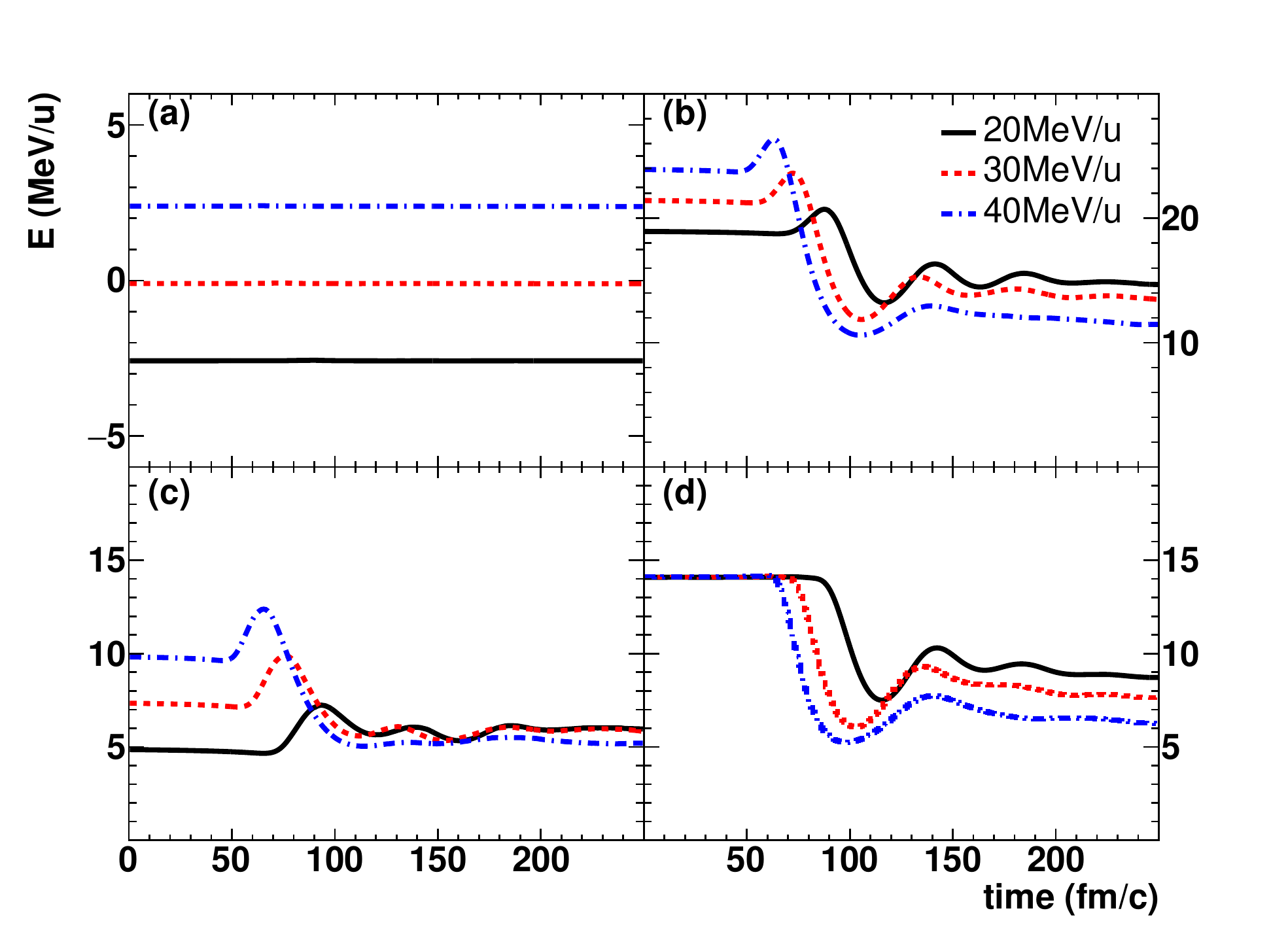}}
\caption{Time evolution of different energy terms:  $\langle \hat{H} \rangle$ (a), $\frac{\langle \hat{p^2} \rangle}{2m}$ (b), $\frac{\langle \hat{p} \rangle^2}{2m}$ (c), $\frac{\langle \hat{p^2} \rangle}{2m}-\frac{\langle \hat{p}\rangle^2}{2m}$ (d) for  $^{14}$N + $^{12}$C at $E/A$ =  20 (black), 30 (red) and 40 (blue) MeV.}
\label{ene_vs_time}
\end{figure}

We plot four terms of energy in Figure \ref{ene_vs_time} from the same reaction to explain why the modified EQMD can reproduce the hard photon production well. The average of binding energy $\langle \hat{H}\rangle$  (a), the average of total kinetic energy $\frac{\langle \hat{p^2}\rangle}{2m}$  (b), the average kinetic energy of the center of the wave packet $\frac{\langle\hat{p}\rangle^2}{2m}$ (c) and the kinetic energy difference from  wave packet $\frac{\langle\hat{p^2}\rangle}{2m}-\frac{\langle\hat{p}\rangle^2}{2m}$ (d) are plotted in different panels. The ground state binding energy is about 7.68 MeV/u for $^{12}$C and about 7.47 MeV/u for $^{14}$N. We can easily calculate the average energy which is about -2.60, -0.11 and 2.38 MeV in $n$-$n$ c.m. frame for this reaction at incident energy $E/A$ = 20, 30 and 40 MeV, respectively.  Figure \ref{ene_vs_time}(a) shows that the results from EQMD model is in a good agreement with theoretical calculations. The lacks of hard photons can not be attributed to the total energy in EQMD model, on the  contrary, it must have another mechanism  leading to  this deficient. In Figure \ref{ene_vs_time}(b), the average kinetic energy  keeps stable before two nuclei touch. The energy will transfer between kinetic energy and potential energy when the collision happens. During the final stage, kinetic energy tends to be stable again. We show the first and second part of kinetic energy from Eq. \ref{eq_h} in Figure \ref{ene_vs_time}(c) and (d), respectively. It is obviously that they are in the same order of amplitude in Fermi energy domain. The conventional two-body collision method only considers the first term during this process. However, it would not give rise to any problem when the  wave packets degree of  freedom  has not been considered. In other words, the kinetic energy can not be stored in the wave packets. However, the width of the wave packets was considered dynamically in the EQMD model, kinetic energy transformation shall take place during the variation of wave packets. Our new method introduced in Sec.~\ref{re-extract_energy}
demonstrates that how the wave packet effects embodied in the collision has been clearly shown.
This is the reason why the new method could give better results.

\begin{figure}[t]
\resizebox{8.6cm}{!}{\includegraphics{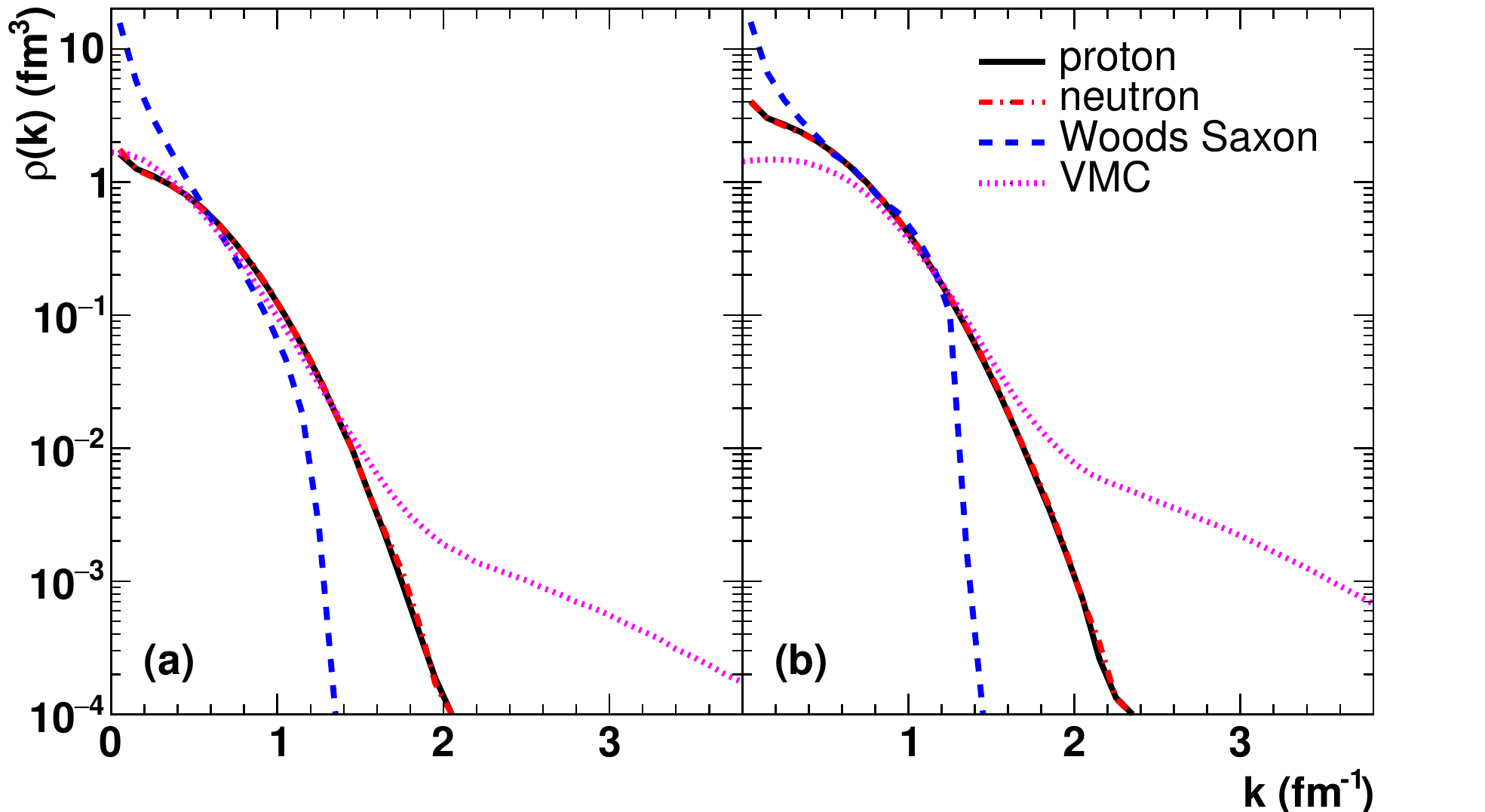}}
\caption{Proton and neutron momentum distribution in $^{4}$He (a) and $^{12}$C (b) after the cooling process. The black solid line and the red dot-dash line represent the proton and neutron calculated by the present EQMD model. The blue dash line represents the proton momentum distribution calculated in the Wood-Saxon distribution under local-density approximation. The pink dot line is taken from the variational Monte Carlo calculation \cite{momenta_distribution}.}
\label{inital_nuclei}
\end{figure}

Because the hard photons should be sensitive to the initial nucleon momentum distribution \cite{ren_photon} in the two colliding nuclei, we also show their momenta distribution after the fraction cooling process. Figure \ref{inital_nuclei} shows momentum distributions of protons (black solid line) and neutrons (red dot-dash line) inside initial nuclei of $^{4}$He (a) and $^{12}$C (b) which was extracted in their momentum representation taking account of zero-point kinetic energy as well as the Wood-Saxon distribution under the local-density approximation and the variational Monte Carlo calculation \cite{momenta_distribution}. Because the effect of short-range correlated (SRC) \cite{Hen, hmt_Libaoan} has not been considered in the EQMD model yet, it is natural that there is no high momentum tail  (HMT) of nucleon distribution as well as no difference in the proton and neutron momentum distribution for symmetric nuclei. Although results from the EQMD model can not match the whole important region from $0.5fm^{-1}$ to $2.5fm^{-1}$, it is obviously better than the Wood-Saxon case. As a result, the momentum distribution after the cooling process of the EQMD model is reasonable as traditional initialization method at least. It is worth to note there is a cutoff in paper of Or Hen et al.\cite{hmt_Libaoan} to simulate the high momentum cutoff obtained from the momentum distribution of deuteron. However, this limit can only be used to limit the initial nuclei, it needs to search another treatment to the sampling method in modified EQMD.
Of course, it will be of interest to  see the effect of SRC in hard-photon production as BUU model \cite{YongGC,ren_photon} by embedding HMT in the EQMD model in near future.

Because  the individual proton-neutron collision is dominant in the incoherent bremsstrahlung process, the angular distribution is nearly isotropic in the $N$-$N$ c.m. frame and the velocity of the $\gamma$-emission source is close to half of the beam rapidity. However, the $N$-$N$ c.m. frame coincides with the $n$-$n$ c.m. frame in the quasisymmetric system. For this reason, it is necessary to check our method in asymmetric systems.

\begin{figure}[htbp]
\resizebox{5cm}{!}{\includegraphics{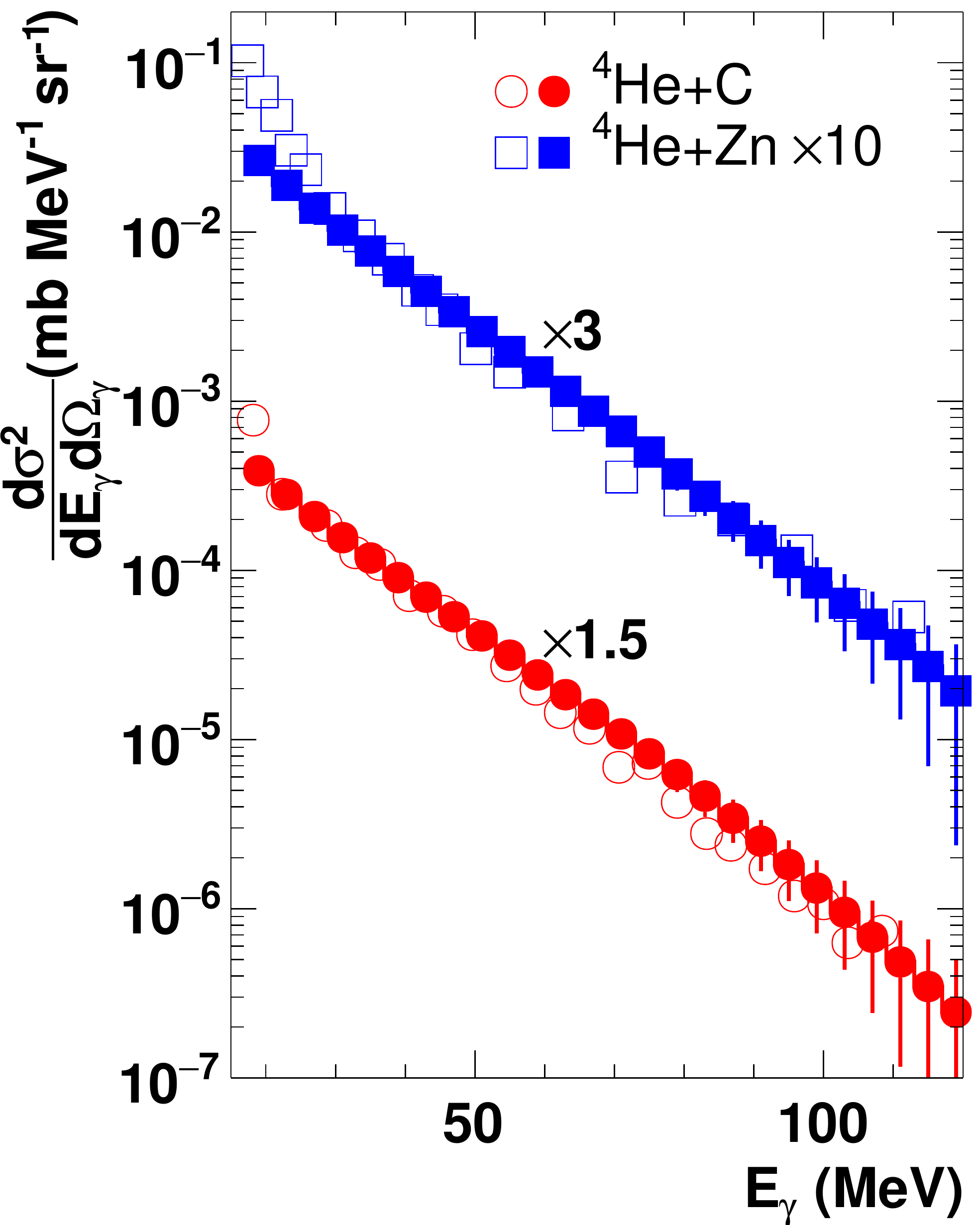}}
\caption{Photon energy spectra with $\theta_{lab}$ around $90^\circ$ ($-0.1$$<$$cos(\theta_{lab})$$<$0.1) calculated by the modified EQMD for reaction $^{4}$He + C (red) and $^{4}$He + Zn (blue) at $E/A$ = 53 MeV are compared to those of experimental data (open markers). The C and Zn target results calculated by the modified EQMD have been multiplied by factors of 1.5 and 3, respectively.}
\label{He_C_Zn}
\end{figure}

\begin{figure}[htbp]
\resizebox{5cm}{!}{\includegraphics{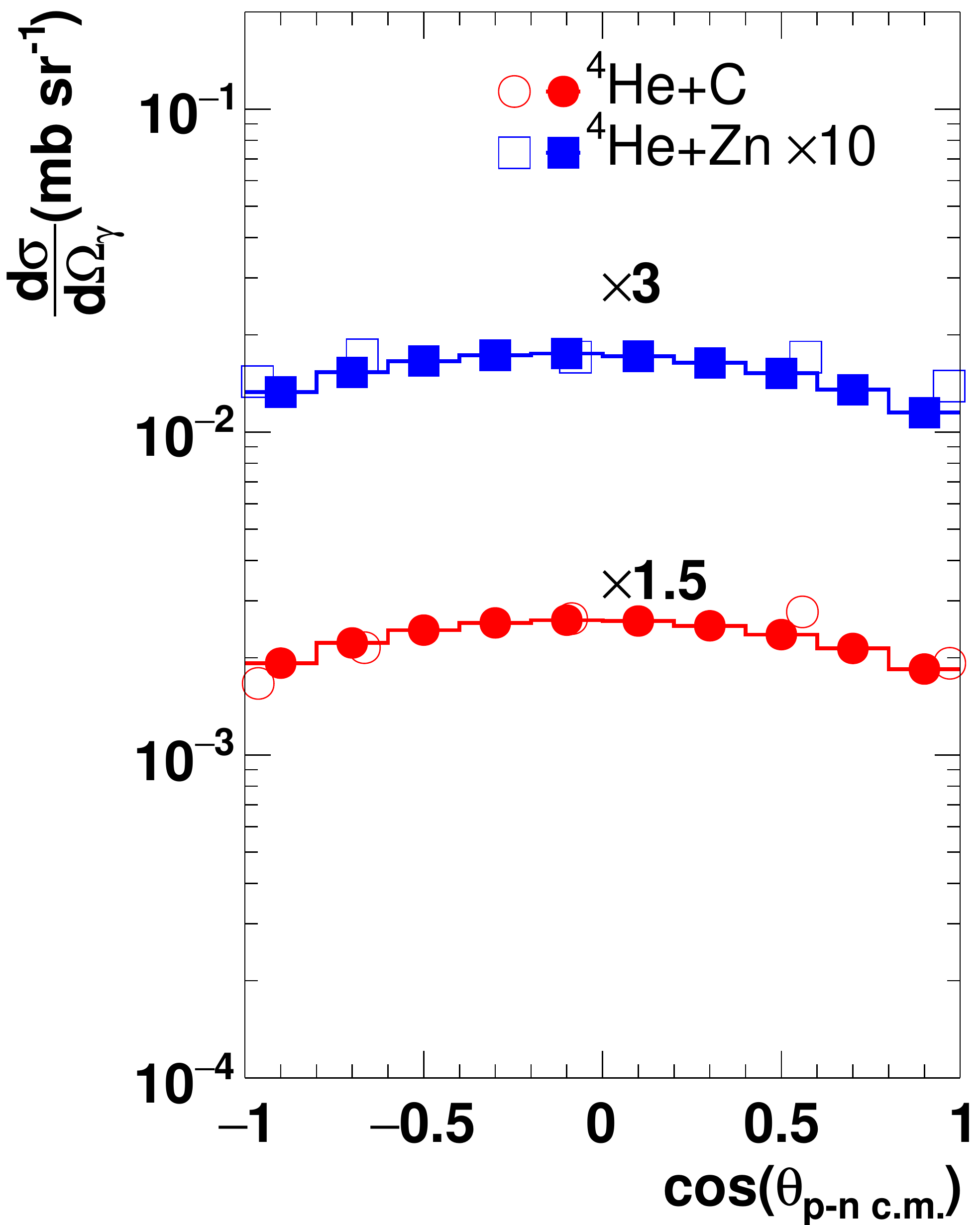}}
\caption{Angular distribution of photons in the proton-neutron center-of-mass  frame with energy $E_{\gamma}>$30 MeV calculated by the modified EQMD for $^{4}$He + C (red) and $^{4}$He + Zn (blue) at  $E/A$ = 53 MeV. The symbols are the same as Fig. 6.}
\label{angle_He_C_Zn}
\end{figure}

\begin{figure}[htbp]
\resizebox{8.6cm}{!}{\includegraphics{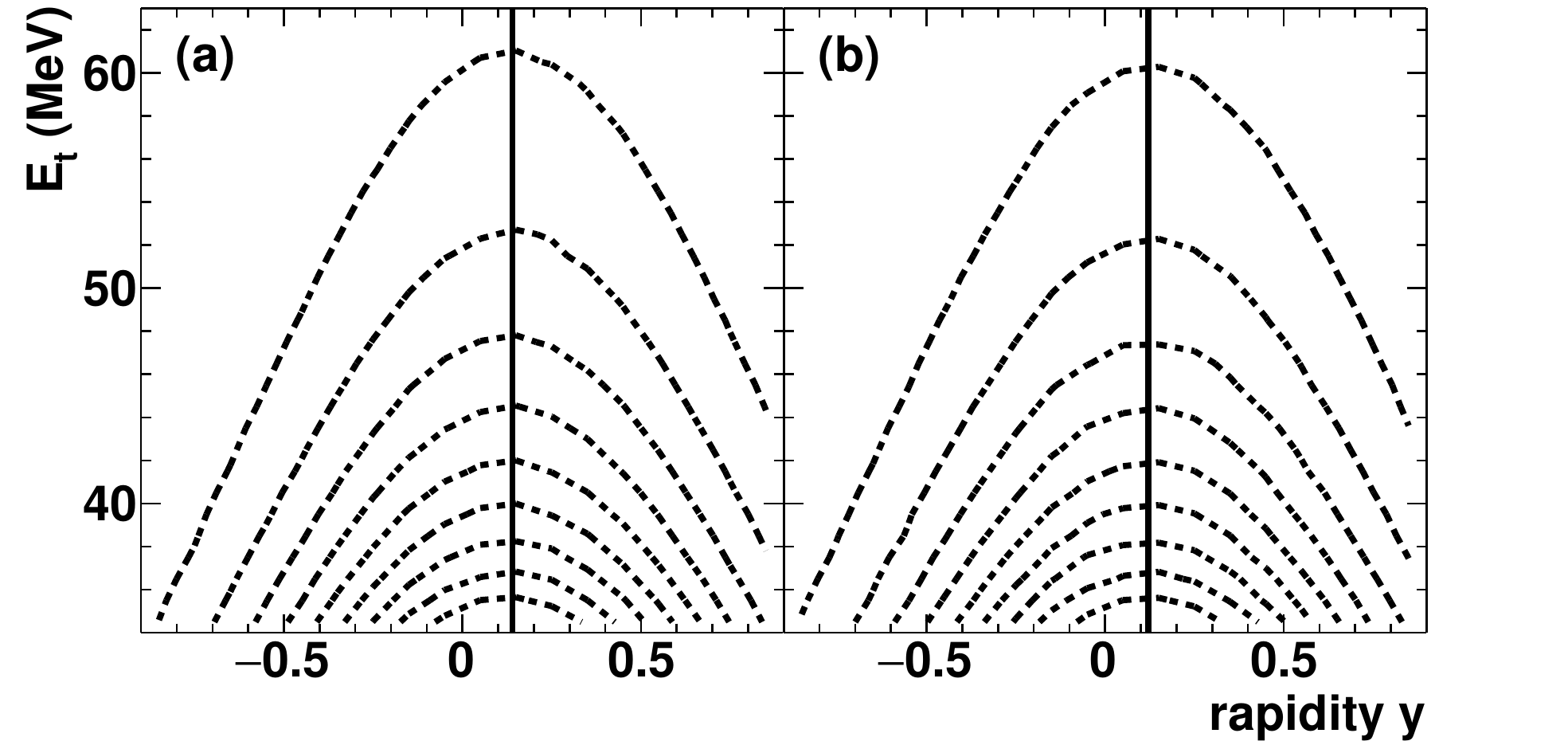}}
\caption{Invariant photon cross section versus rapidity $y$ and transverse energy $E_{t}$ for reactions of $^4$He + $^{12}$C (a) and $^{4}$He + $^{64}$Zn  (b) at $E/A$ = 53 MeV.}
\label{rapidity}
\end{figure}

Figure \ref{He_C_Zn} compares the  energy spectra of hard photons for $^{4}$He + C and $^{4}$He + Zn at incident energy $E/A$ = 53 MeV with our modified EQMD results. We use the $^{12}$C represents $^{nat}$C because its very high abundance. For the same reason, we  consider $^{64}$Zn, $^{66}$Zn, $^{67}$Zn and $^{68}$Zn in the second reaction and their ratio is $0.4863 : 0.2792 : 0.0411 : 0.1934$ for representing natural Zn.
Although the yield of calculations need to be scaled upward by a factor of 1.5 and 3, the inverse slope of the spectra for photon energy above 30 MeV ($E_{\gamma} > 30$MeV) seems in good agreement with the experimental data. This demonstrates that a reasonable collision energy can be realized by the modified EQMD model.

Figure \ref{angle_He_C_Zn} shows the angular distribution of hard photons in the $N$-$N$ c.m. frame for 53 MeV/nucleon $^{4}$He + C (red) and $^{4}$He + Zn (blue). The experimental data show a fairly symmetric distribution about $\theta_{NN} \approx 90^{\circ}$ because the $p$-$n$ collision plays a  dominant role in the incoherent bremsstrahlung process. After being scaled upward by factors of 1.5 and 3, the results calculated by the modified EQMD can well describe the experimental data.
The remaining deviations on yield between our simulation and experimental data might be caused by the uncertainties from the NN cross section, mean field potential, the hard sphere radius or even the deviations from experimental data. It is worth using a more exact sampling method to investigate systematically in the future.

In addition, we extract the velocity of  the $\gamma$-emission source from a contour plot of the invariant $\gamma$-ray emission cross section versus rapidity $y$ and transverse energy $E_t$ \cite{BERTHOLET1987541}. The similar method has also been used to extract the source velocity of $\pi$ \cite{pion_velocity}. Figure \ref{rapidity} shows
the contour of the constant
invariant photon cross section versus rapidity and transverse photon energy (black dotted line) for reactions of $^{4}$He + $^{12}$C in Fig. 8(a) and $^{4}$He + $^{64}$Zn in Fig. 8(b) at $E/A$ = 53 MeV. The shapes of the contour plots are very similar for these two reactions. They are nearly symmetrically distributed about a centroid at rapidity about $y = 0.14$ and $y = 0.12$ (black solid line), respectively. Although those rapidities are less than the half beam rapidity ($\frac{1}{2}y_{beam} = 0.167$) but they are obviously larger than the rapidity of $n$-$n$ c.m. frame which is about 0.084 and 0.020, respectively. It is pointed that the velocity of projectile has some declines during the early stage. Besides, comparing with  light target, heavier target system will slow the source velocity down considerably. This is probably associated with the fact that the second $p$-$n$ collision maybe rise up within the first compression process. In all, both angular distribution in $N$-$N$ c.m. frame and the source velocity confirm that the mean velocity (momentum) of projectile nucleus is not be mistakenly estimated by our method, although we only sample once during a single collision.

\section{Summary}
\label{sec:con}
In this article, we propose a feasible method to improve the performance of the inelastic process, specially for the incoherent $p$-$n$ bremsstrahlung process in the framework of EQMD model.
The energy spectra and angular distributions of direct photons from the reaction of $^{14}$N + $^{12}$C at 20, 30 and 40 MeV/nucleon have been calculated by taking dynamical wave packets effects into account in our modified EQMD model. The magnitude and inverse slope of hard photons from those reactions were reasonably reproduced in comparison with experimental data. In addition, asymmetric reactions of $^{4}$He + C and $^{4}$He + Zn at 53 MeV/nucleon were also simulated in this work. Although their photon yields have to be scaled upward by a factor of 1.5 and 3 in the magnitude, both the inverse slope and  angular distribution shape in the $p$-$n$ c.m. frame, and the velocity of photon source can be correctly reproduced.

The present work uses the direct photon as a sensitive probe of energy and momentum distribution from nuclear reaction during the early stage to test the reliability of the model.
Compared with the original EQMD, those results indicate that the more correct momentum and energy distributions can be obtained in our modified treatment for the EQMD model. The effect of nucleon kinetic energy restored in wave packets is significant around the Fermi energy, and it is necessary to take into account  those contributions in the inelastic process for the models which consider the wave packet width degree of freedom.

\section*{Acknowledgements}
This work was partially supported by the National Natural Science Foundation of China under Contract Nos.11890714, 11421505,  11925502,  U1832129, and 11961141003, the National Key R$\&$D Program of China under Contract No. 2018YFA0404404, the Key Research Program of Frontier Sciences of the CAS under Grant No. QYZDJ-SSW-SLH002, the Youth Innovation Promotion Association CAS under Grant No. 2017309,  and the Strategic Priority Research Program of the CAS under Grant No XDB34030200 and XDB16.

\bibliographystyle{plain}
\bibliography{mylib}

\begin{thebibliography}{55}
\expandafter\ifx\csname natexlab\endcsname\relax\def\natexlab#1{#1}\fi
\expandafter\ifx\csname bibnamefont\endcsname\relax
  \def\bibnamefont#1{#1}\fi
\expandafter\ifx\csname bibfnamefont\endcsname\relax
  \def\bibfnamefont#1{#1}\fi
\expandafter\ifx\csname citenamefont\endcsname\relax
  \def\citenamefont#1{#1}\fi
\expandafter\ifx\csname url\endcsname\relax
  \def\url#1{\texttt{#1}}\fi
\expandafter\ifx\csname urlprefix\endcsname\relax\def\urlprefix{URL }\fi
\providecommand{\bibinfo}[2]{#2}
\providecommand{\eprint}[2][]{\url{#2}}

\bibitem[{\citenamefont{Bonasera et~al.}(2006)\citenamefont{Bonasera,
  Coniglione, and Sapienza}}]{Bonasera}
\bibinfo{author}{\bibfnamefont{A.}~\bibnamefont{Bonasera}},
  \bibinfo{author}{\bibfnamefont{R.}~\bibnamefont{Coniglione}},
  \bibnamefont{and} \bibinfo{author}{\bibfnamefont{P.}~\bibnamefont{Sapienza}},
  \bibinfo{journal}{Eur. Phys. J. A} \textbf{\bibinfo{volume}{30}},
  \bibinfo{pages}{47} (\bibinfo{year}{2006}),
  \urlprefix\url{https://doi.org/10.1140/epja/i2006-10107-8}.

\bibitem[{\citenamefont{Cassing et~al.}(1990)\citenamefont{Cassing, Metag,
  Mosel, and Niita}}]{CassingProduction}
\bibinfo{author}{\bibfnamefont{W.}~\bibnamefont{Cassing}},
  \bibinfo{author}{\bibfnamefont{V.}~\bibnamefont{Metag}},
  \bibinfo{author}{\bibfnamefont{U.}~\bibnamefont{Mosel}}, \bibnamefont{and}
  \bibinfo{author}{\bibfnamefont{K.}~\bibnamefont{Niita}},
  \bibinfo{journal}{Physics Reports} \textbf{\bibinfo{volume}{188}},
  \bibinfo{pages}{363 } (\bibinfo{year}{1990}),
  \urlprefix\url{http://www.sciencedirect.com/science/article/pii/037015739090164W}.

\bibitem[{\citenamefont{Schutz et~al.}(1997)\citenamefont{Schutz, Martinez,
  Marques, Marin, Matulewicz, Ostendorf, Bozek, Delagrange, Diaz, and
  Franke}}]{SchutzHard}
\bibinfo{author}{\bibfnamefont{Y.}~\bibnamefont{Schutz}},
  \bibinfo{author}{\bibfnamefont{G.}~\bibnamefont{Martinez}},
  \bibinfo{author}{\bibfnamefont{F.}~\bibnamefont{Marques}},
  \bibinfo{author}{\bibfnamefont{A.}~\bibnamefont{Marin}},
  \bibinfo{author}{\bibfnamefont{T.}~\bibnamefont{Matulewicz}},
  \bibinfo{author}{\bibfnamefont{R.}~\bibnamefont{Ostendorf}},
  \bibinfo{author}{\bibfnamefont{P.}~\bibnamefont{Bozek}},
  \bibinfo{author}{\bibfnamefont{H.}~\bibnamefont{Delagrange}},
  \bibinfo{author}{\bibfnamefont{J.}~\bibnamefont{Diaz}}, \bibnamefont{and}
  \bibinfo{author}{\bibfnamefont{M.}~\bibnamefont{Franke}},
  \bibinfo{journal}{Nuclear Physics A} \textbf{\bibinfo{volume}{622}},
  \bibinfo{pages}{404} (\bibinfo{year}{1997}),
  \urlprefix\url{https://doi.org/10.1016/S0375-9474(97)00191-7}.

\bibitem[{\citenamefont{Nifenecker and Pinston}(1990)}]{NifeneckerHigh}
\bibinfo{author}{\bibfnamefont{H.}~\bibnamefont{Nifenecker}} \bibnamefont{and}
  \bibinfo{author}{\bibfnamefont{J.~A.} \bibnamefont{Pinston}},
  \bibinfo{journal}{Annual Review of Nuclear and Particle Science}
  \textbf{\bibinfo{volume}{40}}, \bibinfo{pages}{113} (\bibinfo{year}{1990}),
  \urlprefix\url{https://doi.org/10.1146/annurev.ns.40.120190.000553}.

\bibitem[{\citenamefont{Stevenson et~al.}(1986)\citenamefont{Stevenson, Beard,
  Benenson, Clayton, Kashy, Lampis, Morrissey, Samuel, Smith, Tam
  et~al.}}]{N_C}
\bibinfo{author}{\bibfnamefont{J.}~\bibnamefont{Stevenson}},
  \bibinfo{author}{\bibfnamefont{K.~B.} \bibnamefont{Beard}},
  \bibinfo{author}{\bibfnamefont{W.}~\bibnamefont{Benenson}},
  \bibinfo{author}{\bibfnamefont{J.}~\bibnamefont{Clayton}},
  \bibinfo{author}{\bibfnamefont{E.}~\bibnamefont{Kashy}},
  \bibinfo{author}{\bibfnamefont{A.}~\bibnamefont{Lampis}},
  \bibinfo{author}{\bibfnamefont{D.~J.} \bibnamefont{Morrissey}},
  \bibinfo{author}{\bibfnamefont{M.}~\bibnamefont{Samuel}},
  \bibinfo{author}{\bibfnamefont{R.~J.} \bibnamefont{Smith}},
  \bibinfo{author}{\bibfnamefont{C.~L.} \bibnamefont{Tam}},
  \bibnamefont{et~al.}, \bibinfo{journal}{Phys. Rev. Lett.}
  \textbf{\bibinfo{volume}{57}}, \bibinfo{pages}{555} (\bibinfo{year}{1986}),
  \urlprefix\url{https://link.aps.org/doi/10.1103/PhysRevLett.57.555}.

\bibitem[{\citenamefont{Tam et~al.}(1988)\citenamefont{Tam, Stevenson,
  Benenson, Clayton, Chen, Kashy, Lampis, Morrissey, Samuel, Murakami
  et~al.}}]{He_C_Zn}
\bibinfo{author}{\bibfnamefont{C.~L.} \bibnamefont{Tam}},
  \bibinfo{author}{\bibfnamefont{J.}~\bibnamefont{Stevenson}},
  \bibinfo{author}{\bibfnamefont{W.}~\bibnamefont{Benenson}},
  \bibinfo{author}{\bibfnamefont{J.}~\bibnamefont{Clayton}},
  \bibinfo{author}{\bibfnamefont{Y.}~\bibnamefont{Chen}},
  \bibinfo{author}{\bibfnamefont{E.}~\bibnamefont{Kashy}},
  \bibinfo{author}{\bibfnamefont{A.~R.} \bibnamefont{Lampis}},
  \bibinfo{author}{\bibfnamefont{D.~J.} \bibnamefont{Morrissey}},
  \bibinfo{author}{\bibfnamefont{M.}~\bibnamefont{Samuel}},
  \bibinfo{author}{\bibfnamefont{T.~K.} \bibnamefont{Murakami}},
  \bibnamefont{et~al.}, \bibinfo{journal}{Phys. Rev. C}
  \textbf{\bibinfo{volume}{38}}, \bibinfo{pages}{2526} (\bibinfo{year}{1988}),
  \urlprefix\url{https://link.aps.org/doi/10.1103/PhysRevC.38.2526}.

\bibitem[{\citenamefont{Deng and Ma}(2017)}]{DengXG1}
\bibinfo{author}{\bibfnamefont{X.~G.} \bibnamefont{Deng}} \bibnamefont{and}
  \bibinfo{author}{\bibfnamefont{Y.~G.} \bibnamefont{Ma}},
  \bibinfo{journal}{Nucl. Sci. Tech.} \textbf{\bibinfo{volume}{28}},
  \bibinfo{pages}{182} (\bibinfo{year}{2017}),
  \urlprefix\url{https://doi.org/10.1007/s41365-017-0337-1}.

\bibitem[{\citenamefont{Deng and Ma}(2018)}]{DengXG2}
\bibinfo{author}{\bibfnamefont{X.~G.} \bibnamefont{Deng}} \bibnamefont{and}
  \bibinfo{author}{\bibfnamefont{Y.~G.} \bibnamefont{Ma}},
  \bibinfo{journal}{Eur. Phys. J. A} \textbf{\bibinfo{volume}{54}},
  \bibinfo{pages}{204} (\bibinfo{year}{2018}),
  \urlprefix\url{https://doi.org/10.1140/epja/i2018-12635-x}.

\bibitem[{\citenamefont{Kafexhiu}(2016)}]{Kafexhiu}
\bibinfo{author}{\bibfnamefont{E.}~\bibnamefont{Kafexhiu}},
  \bibinfo{journal}{Phys. Rev. C} \textbf{\bibinfo{volume}{94}},
  \bibinfo{pages}{064603} (\bibinfo{year}{2016}),
  \urlprefix\url{https://link.aps.org/doi/10.1103/PhysRevC.94.064603}.

\bibitem[{\citenamefont{Ko et~al.}(1985)\citenamefont{Ko, Bertsch, and
  Aichelin}}]{incoherent}
\bibinfo{author}{\bibfnamefont{C.~M.} \bibnamefont{Ko}},
  \bibinfo{author}{\bibfnamefont{G.}~\bibnamefont{Bertsch}}, \bibnamefont{and}
  \bibinfo{author}{\bibfnamefont{J.}~\bibnamefont{Aichelin}},
  \bibinfo{journal}{Phys. Rev. C} \textbf{\bibinfo{volume}{31}},
  \bibinfo{pages}{2324} (\bibinfo{year}{1985}),
  \urlprefix\url{https://link.aps.org/doi/10.1103/PhysRevC.31.2324}.

\bibitem[{\citenamefont{Bauer et~al.}(1986{\natexlab{a}})\citenamefont{Bauer,
  Bertsch, Cassing, and Mosel}}]{PhysRevC.34.2127}
\bibinfo{author}{\bibfnamefont{W.}~\bibnamefont{Bauer}},
  \bibinfo{author}{\bibfnamefont{G.~F.} \bibnamefont{Bertsch}},
  \bibinfo{author}{\bibfnamefont{W.}~\bibnamefont{Cassing}}, \bibnamefont{and}
  \bibinfo{author}{\bibfnamefont{U.}~\bibnamefont{Mosel}},
  \bibinfo{journal}{Phys. Rev. C} \textbf{\bibinfo{volume}{34}},
  \bibinfo{pages}{2127} (\bibinfo{year}{1986}{\natexlab{a}}),
  \urlprefix\url{https://link.aps.org/doi/10.1103/PhysRevC.34.2127}.

\bibitem[{\citenamefont{Vasak et~al.}(1985)\citenamefont{Vasak, M\"uller, and
  Greiner}}]{Vasak_1985}
\bibinfo{author}{\bibfnamefont{D.}~\bibnamefont{Vasak}},
  \bibinfo{author}{\bibfnamefont{B.}~\bibnamefont{M\"uller}}, \bibnamefont{and}
  \bibinfo{author}{\bibfnamefont{W.}~\bibnamefont{Greiner}},
  \bibinfo{journal}{Journal of Physics G: Nuclear Physics}
  \textbf{\bibinfo{volume}{11}}, \bibinfo{pages}{1309} (\bibinfo{year}{1985}),
  \urlprefix\url{https://doi.org/10.1088%2F0305-4616%2F11%2F12%2F009}.

\bibitem[{\citenamefont{Vasak}(1986)}]{VASAK1986276}
\bibinfo{author}{\bibfnamefont{D.}~\bibnamefont{Vasak}},
  \bibinfo{journal}{Physics Letters B} \textbf{\bibinfo{volume}{176}},
  \bibinfo{pages}{276 } (\bibinfo{year}{1986}),
  \urlprefix\url{http://www.sciencedirect.com/science/article/pii/0370269386901632}.

\bibitem[{\citenamefont{Bauer et~al.}(1986{\natexlab{b}})\citenamefont{Bauer,
  Cassing, Mosel, Tohyama, and Cusson}}]{BAUER1986159}
\bibinfo{author}{\bibfnamefont{W.}~\bibnamefont{Bauer}},
  \bibinfo{author}{\bibfnamefont{W.}~\bibnamefont{Cassing}},
  \bibinfo{author}{\bibfnamefont{U.}~\bibnamefont{Mosel}},
  \bibinfo{author}{\bibfnamefont{M.}~\bibnamefont{Tohyama}}, \bibnamefont{and}
  \bibinfo{author}{\bibfnamefont{R.}~\bibnamefont{Cusson}},
  \bibinfo{journal}{Nuclear Physics A} \textbf{\bibinfo{volume}{456}},
  \bibinfo{pages}{159 } (\bibinfo{year}{1986}{\natexlab{b}}),
  \urlprefix\url{http://www.sciencedirect.com/science/article/pii/0375947486903714}.

\bibitem[{\citenamefont{Bertholet et~al.}(1987)\citenamefont{Bertholet, Njock,
  Maurel, Monnand, Nifenecker, Perrin, Pinston, Schussler, Barneoud, Guet
  et~al.}}]{BERTHOLET1987541}
\bibinfo{author}{\bibfnamefont{R.}~\bibnamefont{Bertholet}},
  \bibinfo{author}{\bibfnamefont{M.~K.} \bibnamefont{Njock}},
  \bibinfo{author}{\bibfnamefont{M.}~\bibnamefont{Maurel}},
  \bibinfo{author}{\bibfnamefont{E.}~\bibnamefont{Monnand}},
  \bibinfo{author}{\bibfnamefont{H.}~\bibnamefont{Nifenecker}},
  \bibinfo{author}{\bibfnamefont{P.}~\bibnamefont{Perrin}},
  \bibinfo{author}{\bibfnamefont{J.}~\bibnamefont{Pinston}},
  \bibinfo{author}{\bibfnamefont{F.}~\bibnamefont{Schussler}},
  \bibinfo{author}{\bibfnamefont{D.}~\bibnamefont{Barneoud}},
  \bibinfo{author}{\bibfnamefont{C.}~\bibnamefont{Guet}}, \bibnamefont{et~al.},
  \bibinfo{journal}{Nuclear Physics A} \textbf{\bibinfo{volume}{474}},
  \bibinfo{pages}{541 } (\bibinfo{year}{1987}),
  \urlprefix\url{http://www.sciencedirect.com/science/article/pii/0375947487906300}.

\bibitem[{\citenamefont{Biro et~al.}(1987)\citenamefont{Biro, Niita, Paoli,
  Bauer, Cassing, and Mosel}}]{BIRO1987579}
\bibinfo{author}{\bibfnamefont{T.}~\bibnamefont{Biro}},
  \bibinfo{author}{\bibfnamefont{K.}~\bibnamefont{Niita}},
  \bibinfo{author}{\bibfnamefont{A.~D.} \bibnamefont{Paoli}},
  \bibinfo{author}{\bibfnamefont{W.}~\bibnamefont{Bauer}},
  \bibinfo{author}{\bibfnamefont{W.}~\bibnamefont{Cassing}}, \bibnamefont{and}
  \bibinfo{author}{\bibfnamefont{U.}~\bibnamefont{Mosel}},
  \bibinfo{journal}{Nuclear Physics A} \textbf{\bibinfo{volume}{475}},
  \bibinfo{pages}{579 } (\bibinfo{year}{1987}),
  \urlprefix\url{http://www.sciencedirect.com/science/article/pii/0375947487900807}.

\bibitem[{\citenamefont{Liu et~al.}(2008)\citenamefont{Liu, Ma, Cai, Fang,
  Shen, Tian, and Wang}}]{LiuGH2008}
\bibinfo{author}{\bibfnamefont{G.~H.} \bibnamefont{Liu}},
  \bibinfo{author}{\bibfnamefont{Y.~G.} \bibnamefont{Ma}},
  \bibinfo{author}{\bibfnamefont{X.~Z.} \bibnamefont{Cai}},
  \bibinfo{author}{\bibfnamefont{D.~Q.} \bibnamefont{Fang}},
  \bibinfo{author}{\bibfnamefont{W.~Q.} \bibnamefont{Shen}},
  \bibinfo{author}{\bibfnamefont{W.~D.} \bibnamefont{Tian}}, \bibnamefont{and}
  \bibinfo{author}{\bibfnamefont{K.}~\bibnamefont{Wang}},
  \bibinfo{journal}{Phys. Lett. B} \textbf{\bibinfo{volume}{663}},
  \bibinfo{pages}{312} (\bibinfo{year}{2008}),
  \urlprefix\url{https://doi.org/10.1016/j.physletb.2008.04.037}.

\bibitem[{\citenamefont{Ma et~al.}(2012)\citenamefont{Ma, Liu, Cai, Fang, Guo,
  Shen, Tian, and Wang}}]{MaYG2012}
\bibinfo{author}{\bibfnamefont{Y.~G.} \bibnamefont{Ma}},
  \bibinfo{author}{\bibfnamefont{G.~H.} \bibnamefont{Liu}},
  \bibinfo{author}{\bibfnamefont{X.~Z.} \bibnamefont{Cai}},
  \bibinfo{author}{\bibfnamefont{D.~Q.} \bibnamefont{Fang}},
  \bibinfo{author}{\bibfnamefont{W.}~\bibnamefont{Guo}},
  \bibinfo{author}{\bibfnamefont{W.}~\bibnamefont{Shen}},
  \bibinfo{author}{\bibfnamefont{W.~D.} \bibnamefont{Tian}}, \bibnamefont{and}
  \bibinfo{author}{\bibfnamefont{H.~W.} \bibnamefont{Wang}},
  \bibinfo{journal}{Phys. Rev. C} \textbf{\bibinfo{volume}{85}},
  \bibinfo{pages}{024618} (\bibinfo{year}{2012}),
  \urlprefix\url{https://doi.org/10.1103/PhysRevC.85.024618}.

\bibitem[{\citenamefont{Yong and Li}(2017)}]{YongGC}
\bibinfo{author}{\bibfnamefont{G.~C.} \bibnamefont{Yong}} \bibnamefont{and}
  \bibinfo{author}{\bibfnamefont{B.~A.} \bibnamefont{Li}},
  \bibinfo{journal}{Phys. Rev. C} \textbf{\bibinfo{volume}{96}},
  \bibinfo{pages}{064614} (\bibinfo{year}{2017}),
  \urlprefix\url{https://doi.org/10.1103/PhysRevC.96.064614}.

\bibitem[{\citenamefont{Tam et~al.}(1989)\citenamefont{Tam, Stevenson,
  Benenson, Clayton, Chen, Kashy, Lampis, Morrissey, Samuel, Murakami
  et~al.}}]{Tam_incoherent}
\bibinfo{author}{\bibfnamefont{C.~L.} \bibnamefont{Tam}},
  \bibinfo{author}{\bibfnamefont{J.}~\bibnamefont{Stevenson}},
  \bibinfo{author}{\bibfnamefont{W.}~\bibnamefont{Benenson}},
  \bibinfo{author}{\bibfnamefont{J.}~\bibnamefont{Clayton}},
  \bibinfo{author}{\bibfnamefont{Y.}~\bibnamefont{Chen}},
  \bibinfo{author}{\bibfnamefont{E.}~\bibnamefont{Kashy}},
  \bibinfo{author}{\bibfnamefont{A.~R.} \bibnamefont{Lampis}},
  \bibinfo{author}{\bibfnamefont{D.~J.} \bibnamefont{Morrissey}},
  \bibinfo{author}{\bibfnamefont{M.}~\bibnamefont{Samuel}},
  \bibinfo{author}{\bibfnamefont{T.~K.} \bibnamefont{Murakami}},
  \bibnamefont{et~al.}, \bibinfo{journal}{Phys. Rev. C}
  \textbf{\bibinfo{volume}{39}}, \bibinfo{pages}{1371} (\bibinfo{year}{1989}),
  \urlprefix\url{https://link.aps.org/doi/10.1103/PhysRevC.39.1371}.

\bibitem[{\citenamefont{Maruyama et~al.}(1996)\citenamefont{Maruyama, Niita,
  and Iwamoto}}]{EQMD}
\bibinfo{author}{\bibfnamefont{T.}~\bibnamefont{Maruyama}},
  \bibinfo{author}{\bibfnamefont{K.}~\bibnamefont{Niita}}, \bibnamefont{and}
  \bibinfo{author}{\bibfnamefont{A.}~\bibnamefont{Iwamoto}},
  \bibinfo{journal}{Phys. Rev. C} \textbf{\bibinfo{volume}{53}},
  \bibinfo{pages}{297} (\bibinfo{year}{1996}),
  \urlprefix\url{https://link.aps.org/doi/10.1103/PhysRevC.53.297}.

\bibitem[{\citenamefont{Wada et~al.}(1998)\citenamefont{Wada, Hagel, Cibor, Li,
  Marie, Shen, Zhao, Natowitz, and Ono}}]{hagel}
\bibinfo{author}{\bibfnamefont{R.}~\bibnamefont{Wada}},
  \bibinfo{author}{\bibfnamefont{K.~.} \bibnamefont{Hagel}},
  \bibinfo{author}{\bibfnamefont{J.}~\bibnamefont{Cibor}},
  \bibinfo{author}{\bibfnamefont{J.}~\bibnamefont{Li}},
  \bibinfo{author}{\bibfnamefont{N.}~\bibnamefont{Marie}},
  \bibinfo{author}{\bibfnamefont{W.}~\bibnamefont{Shen}},
  \bibinfo{author}{\bibfnamefont{Y.}~\bibnamefont{Zhao}},
  \bibinfo{author}{\bibfnamefont{J.~B.} \bibnamefont{Natowitz}},
  \bibnamefont{and} \bibinfo{author}{\bibfnamefont{A.}~\bibnamefont{Ono}},
  \bibinfo{journal}{Phys. Lett. B} \textbf{\bibinfo{volume}{422}},
  \bibinfo{pages}{6} (\bibinfo{year}{1998}),
  \urlprefix\url{https://doi.org/10.1016/S0370-2693(98)00033-1}.

\bibitem[{\citenamefont{He et~al.}(2014)\citenamefont{He, Ma, Cao, Cai, and
  Zhang}}]{He_wb_gdr_prl}
\bibinfo{author}{\bibfnamefont{W.~B.} \bibnamefont{He}},
  \bibinfo{author}{\bibfnamefont{Y.~G.} \bibnamefont{Ma}},
  \bibinfo{author}{\bibfnamefont{X.~G.} \bibnamefont{Cao}},
  \bibinfo{author}{\bibfnamefont{X.~Z.} \bibnamefont{Cai}}, \bibnamefont{and}
  \bibinfo{author}{\bibfnamefont{G.~Q.} \bibnamefont{Zhang}},
  \bibinfo{journal}{Phys. Rev. Lett.} \textbf{\bibinfo{volume}{113}},
  \bibinfo{pages}{032506} (\bibinfo{year}{2014}),
  \urlprefix\url{https://link.aps.org/doi/10.1103/PhysRevLett.113.032506}.

\bibitem[{\citenamefont{He et~al.}(2016)\citenamefont{He, Ma, Cao, Cai, and
  Zhang}}]{HeWB2}
\bibinfo{author}{\bibfnamefont{W.~B.} \bibnamefont{He}},
  \bibinfo{author}{\bibfnamefont{Y.~G.} \bibnamefont{Ma}},
  \bibinfo{author}{\bibfnamefont{X.~G.} \bibnamefont{Cao}},
  \bibinfo{author}{\bibfnamefont{X.~Z.} \bibnamefont{Cai}}, \bibnamefont{and}
  \bibinfo{author}{\bibfnamefont{G.~Q.} \bibnamefont{Zhang}},
  \bibinfo{journal}{Phys. Rev. C} \textbf{\bibinfo{volume}{94}},
  \bibinfo{pages}{014301} (\bibinfo{year}{2016}),
  \urlprefix\url{https://doi.org/10.1103/PhysRevC.94.014301}.

\bibitem[{\citenamefont{Wang et~al.}(2017{\natexlab{a}})\citenamefont{Wang, Ma,
  Cao, Kong, and Ma}}]{WangSS}
\bibinfo{author}{\bibfnamefont{S.~S.} \bibnamefont{Wang}},
  \bibinfo{author}{\bibfnamefont{Y.~G.} \bibnamefont{Ma}},
  \bibinfo{author}{\bibfnamefont{W.~B.} \bibnamefont{Cao},
  \bibfnamefont{X.~G.and~He}}, \bibinfo{author}{\bibfnamefont{H.~Y.}
  \bibnamefont{Kong}}, \bibnamefont{and} \bibinfo{author}{\bibfnamefont{C.~W.}
  \bibnamefont{Ma}}, \bibinfo{journal}{Phys. Rev. C}
  \textbf{\bibinfo{volume}{95}}, \bibinfo{pages}{054615}
  (\bibinfo{year}{2017}{\natexlab{a}}),
  \urlprefix\url{https://doi.org/10.1103/PhysRevC.95.054615}.

\bibitem[{\citenamefont{Wang et~al.}(2017{\natexlab{b}})\citenamefont{Wang, Ma,
  Zhang, Cao, He, and Shen}}]{WangK}
\bibinfo{author}{\bibfnamefont{K.}~\bibnamefont{Wang}},
  \bibinfo{author}{\bibfnamefont{Y.~G.} \bibnamefont{Ma}},
  \bibinfo{author}{\bibfnamefont{G.~Q.} \bibnamefont{Zhang}},
  \bibinfo{author}{\bibfnamefont{X.~G.} \bibnamefont{Cao}},
  \bibinfo{author}{\bibfnamefont{W.~B.} \bibnamefont{He}}, \bibnamefont{and}
  \bibinfo{author}{\bibfnamefont{W.~Q.} \bibnamefont{Shen}},
  \bibinfo{journal}{Phys. Rev. C} \textbf{\bibinfo{volume}{95}},
  \bibinfo{pages}{014608} (\bibinfo{year}{2017}{\natexlab{b}}),
  \urlprefix\url{https://doi.org/10.1103/PhysRevC.95.014608}.

\bibitem[{\citenamefont{Huang et~al.}(2017)\citenamefont{Huang, Ma, and
  He}}]{Huang_bs}
\bibinfo{author}{\bibfnamefont{B.~S.} \bibnamefont{Huang}},
  \bibinfo{author}{\bibfnamefont{Y.~G.} \bibnamefont{Ma}}, \bibnamefont{and}
  \bibinfo{author}{\bibfnamefont{W.~B.} \bibnamefont{He}},
  \bibinfo{journal}{Phys. Rev. C} \textbf{\bibinfo{volume}{95}},
  \bibinfo{pages}{034606} (\bibinfo{year}{2017}),
  \urlprefix\url{https://link.aps.org/doi/10.1103/PhysRevC.95.034606}.

\bibitem[{\citenamefont{Guo et~al.}(2019)\citenamefont{Guo, Ma, An, and
  Huang}}]{guo_cc}
\bibinfo{author}{\bibfnamefont{C.-C.} \bibnamefont{Guo}},
  \bibinfo{author}{\bibfnamefont{Y.-G.} \bibnamefont{Ma}},
  \bibinfo{author}{\bibfnamefont{Z.-D.} \bibnamefont{An}}, \bibnamefont{and}
  \bibinfo{author}{\bibfnamefont{B.-S.} \bibnamefont{Huang}},
  \bibinfo{journal}{Phys. Rev. C} \textbf{\bibinfo{volume}{99}},
  \bibinfo{pages}{044607} (\bibinfo{year}{2019}),
  \urlprefix\url{https://link.aps.org/doi/10.1103/PhysRevC.99.044607}.

\bibitem[{\citenamefont{Guo et~al.}(2017)\citenamefont{Guo, Ma, and He~et
  al.}}]{GuoCQ}
\bibinfo{author}{\bibfnamefont{C.~Q.} \bibnamefont{Guo}},
  \bibinfo{author}{\bibfnamefont{Y.~G.} \bibnamefont{Ma}}, \bibnamefont{and}
  \bibinfo{author}{\bibfnamefont{W.~B.} \bibnamefont{He~et al.}},
  \bibinfo{journal}{Phys. Rev. C} \textbf{\bibinfo{volume}{95}},
  \bibinfo{pages}{054622} (\bibinfo{year}{2017}),
  \urlprefix\url{https://doi.org/10.1103/PhysRevC.95.054622}.

\bibitem[{\citenamefont{Aichelin}(1991)}]{Aichelin}
\bibinfo{author}{\bibfnamefont{J.}~\bibnamefont{Aichelin}},
  \bibinfo{journal}{Phys. Rep.} \textbf{\bibinfo{volume}{202}},
  \bibinfo{pages}{233} (\bibinfo{year}{1991}),
  \urlprefix\url{https://doi.org/10.1016/0370-1573(91)90094-3}.

\bibitem[{\citenamefont{Hartnack et~al.}(1998)\citenamefont{Hartnack, Puri,
  Aichelin, Konopka, Bass, Stoecker, and Greiner}}]{Hartnack1998Modelling}
\bibinfo{author}{\bibfnamefont{C.}~\bibnamefont{Hartnack}},
  \bibinfo{author}{\bibfnamefont{R.~K.} \bibnamefont{Puri}},
  \bibinfo{author}{\bibfnamefont{J.}~\bibnamefont{Aichelin}},
  \bibinfo{author}{\bibfnamefont{J.}~\bibnamefont{Konopka}},
  \bibinfo{author}{\bibfnamefont{S.~A.} \bibnamefont{Bass}},
  \bibinfo{author}{\bibfnamefont{H.}~\bibnamefont{Stoecker}}, \bibnamefont{and}
  \bibinfo{author}{\bibfnamefont{W.}~\bibnamefont{Greiner}},
  \bibinfo{journal}{The European Physical Journal A - Hadrons and Nuclei}
  \textbf{\bibinfo{volume}{1}}, \bibinfo{pages}{151} (\bibinfo{year}{1998}),
  \urlprefix\url{https://doi.org/10.1007/s100500050045}.

\bibitem[{\citenamefont{Yan and Li}(2019)}]{Yan1}
\bibinfo{author}{\bibfnamefont{T.~Z.} \bibnamefont{Yan}} \bibnamefont{and}
  \bibinfo{author}{\bibfnamefont{S.}~\bibnamefont{Li}}, \bibinfo{journal}{Nucl.
  Sci. Tech.} \textbf{\bibinfo{volume}{30}}, \bibinfo{pages}{43}
  (\bibinfo{year}{2019}),
  \urlprefix\url{https://doi.org/10.1007/s41365-019-0572-8}.

\bibitem[{\citenamefont{Wang et~al.}(2019)\citenamefont{Wang, Ma, and
  Zhang}}]{WangTT}
\bibinfo{author}{\bibfnamefont{T.-T.} \bibnamefont{Wang}},
  \bibinfo{author}{\bibfnamefont{Y.-G.} \bibnamefont{Ma}}, \bibnamefont{and}
  \bibinfo{author}{\bibfnamefont{Z.-Q.} \bibnamefont{Zhang}},
  \bibinfo{journal}{Phys. Rev. C} \textbf{\bibinfo{volume}{99}},
  \bibinfo{pages}{054626} (\bibinfo{year}{2019}),
  \urlprefix\url{https://doi.org/10.1103/PhysRevC.99.054626}.

\bibitem[{\citenamefont{Yan et~al.}(2019)\citenamefont{Yan, Li, Wang, and
  Xie}}]{Yan2}
\bibinfo{author}{\bibfnamefont{T.~Z.} \bibnamefont{Yan}},
  \bibinfo{author}{\bibfnamefont{S.}~\bibnamefont{Li}},
  \bibinfo{author}{\bibfnamefont{Y.~N.} \bibnamefont{Wang}}, \bibnamefont{and}
  \bibinfo{author}{\bibfnamefont{F.}~\bibnamefont{Xie}},
  \bibinfo{journal}{Nucl. Sci. Tech.} \textbf{\bibinfo{volume}{30}},
  \bibinfo{pages}{15} (\bibinfo{year}{2019}),
  \urlprefix\url{https://doi.org/10.1007/s41365-018-0534-6}.

\bibitem[{\citenamefont{Li et~al.}(2018)\citenamefont{Li, Wang, Li, and
  Zhang}}]{LiPC}
\bibinfo{author}{\bibfnamefont{P.~C.} \bibnamefont{Li}},
  \bibinfo{author}{\bibfnamefont{Y.~J.} \bibnamefont{Wang}},
  \bibinfo{author}{\bibfnamefont{Q.~F.} \bibnamefont{Li}}, \bibnamefont{and}
  \bibinfo{author}{\bibfnamefont{H.~F.} \bibnamefont{Zhang}},
  \bibinfo{journal}{Nucl. Sci. Tech.} \textbf{\bibinfo{volume}{29}},
  \bibinfo{pages}{177} (\bibinfo{year}{2018}),
  \urlprefix\url{https://doi.org/10.1007/s41365-018-0510-1}.

\bibitem[{\citenamefont{Ono et~al.}(2019)\citenamefont{Ono, Xu, and Colonna~et
  al.}}]{Ono}
\bibinfo{author}{\bibfnamefont{A.}~\bibnamefont{Ono}},
  \bibinfo{author}{\bibfnamefont{J.}~\bibnamefont{Xu}}, \bibnamefont{and}
  \bibinfo{author}{\bibfnamefont{M.}~\bibnamefont{Colonna~et al.}},
  \bibinfo{journal}{Phys. Rev. C} \textbf{\bibinfo{volume}{100}},
  \bibinfo{pages}{044617} (\bibinfo{year}{2019}),
  \urlprefix\url{https://doi.org/10.1103/PhysRevC.100.044617}.

\bibitem[{\citenamefont{Zhang et~al.}(2018)\citenamefont{Zhang, Fang, and
  Ma}}]{ZhangZF}
\bibinfo{author}{\bibfnamefont{Z.~F.} \bibnamefont{Zhang}},
  \bibinfo{author}{\bibfnamefont{D.~Q.} \bibnamefont{Fang}}, \bibnamefont{and}
  \bibinfo{author}{\bibfnamefont{Y.~G.} \bibnamefont{Ma}},
  \bibinfo{journal}{Nucl. Sci. Tech.} \textbf{\bibinfo{volume}{29}},
  \bibinfo{pages}{78} (\bibinfo{year}{2018}),
  \urlprefix\url{https://doi.org/10.1007/s41365-018-0427-8}.

\bibitem[{\citenamefont{Feng}(2018)}]{FengZQ}
\bibinfo{author}{\bibfnamefont{Z.~Q.} \bibnamefont{Feng}},
  \bibinfo{journal}{Nucl. Sci. Tech.} \textbf{\bibinfo{volume}{29}},
  \bibinfo{pages}{40} (\bibinfo{year}{2018}),
  \urlprefix\url{https://doi.org/10.1007/s41365-018-0379-z}.

\bibitem[{\citenamefont{Sood et~al.}(2019)\citenamefont{Sood, Kumar, Sharma,
  and Puri}}]{Sood}
\bibinfo{author}{\bibfnamefont{S.}~\bibnamefont{Sood}},
  \bibinfo{author}{\bibfnamefont{R.}~\bibnamefont{Kumar}},
  \bibinfo{author}{\bibfnamefont{A.}~\bibnamefont{Sharma}}, \bibnamefont{and}
  \bibinfo{author}{\bibfnamefont{R.~K.} \bibnamefont{Puri}},
  \bibinfo{journal}{Phys. Rev. C} \textbf{\bibinfo{volume}{99}},
  \bibinfo{pages}{054612} (\bibinfo{year}{2019}),
  \urlprefix\url{https://doi.org/10.1103/PhysRevC.99.054612}.

\bibitem[{\citenamefont{Desai et~al.}(2019)\citenamefont{Desai, Loveland, and
  McCaleb~et al.}}]{Desai}
\bibinfo{author}{\bibfnamefont{V.~V.} \bibnamefont{Desai}},
  \bibinfo{author}{\bibfnamefont{W.}~\bibnamefont{Loveland}}, \bibnamefont{and}
  \bibinfo{author}{\bibfnamefont{K.}~\bibnamefont{McCaleb~et al.}},
  \bibinfo{journal}{Phys. Rev. C} \textbf{\bibinfo{volume}{99}},
  \bibinfo{pages}{044604} (\bibinfo{year}{2019}),
  \urlprefix\url{https://doi.org/10.1103/PhysRevC.99.044604}.

\bibitem[{\citenamefont{Liu et~al.}(2019)\citenamefont{Liu, Ma, and
  Fang}}]{LiuHL}
\bibinfo{author}{\bibfnamefont{H.~L.} \bibnamefont{Liu}},
  \bibinfo{author}{\bibfnamefont{Y.~G.} \bibnamefont{Ma}}, \bibnamefont{and}
  \bibinfo{author}{\bibfnamefont{D.~Q.} \bibnamefont{Fang}},
  \bibinfo{journal}{Phys. Rev. C} \textbf{\bibinfo{volume}{99}},
  \bibinfo{pages}{054614} (\bibinfo{year}{2019}),
  \urlprefix\url{https://doi.org/10.1103/PhysRevC.99.054614}.

\bibitem[{\citenamefont{Metag}(1988)}]{METAG1988483}
\bibinfo{author}{\bibfnamefont{V.}~\bibnamefont{Metag}},
  \bibinfo{journal}{Nuclear Physics A} \textbf{\bibinfo{volume}{488}},
  \bibinfo{pages}{483 } (\bibinfo{year}{1988}),
  \urlprefix\url{http://www.sciencedirect.com/science/article/pii/0375947488902849}.

\bibitem[{\citenamefont{d'Enterria et~al.}(2002)\citenamefont{d'Enterria,
  Martinez, Aphecetche, Delagrange, Fernandez, Lohner, Ortega, Ostendorf,
  Schutz, and Wilschut}}]{DENTERRIA200227}
\bibinfo{author}{\bibfnamefont{D.}~\bibnamefont{d'Enterria}},
  \bibinfo{author}{\bibfnamefont{G.}~\bibnamefont{Martinez}},
  \bibinfo{author}{\bibfnamefont{L.}~\bibnamefont{Aphecetche}},
  \bibinfo{author}{\bibfnamefont{H.}~\bibnamefont{Delagrange}},
  \bibinfo{author}{\bibfnamefont{F.}~\bibnamefont{Fernandez}},
  \bibinfo{author}{\bibfnamefont{H.}~\bibnamefont{Lohner}},
  \bibinfo{author}{\bibfnamefont{R.}~\bibnamefont{Ortega}},
  \bibinfo{author}{\bibfnamefont{R.}~\bibnamefont{Ostendorf}},
  \bibinfo{author}{\bibfnamefont{Y.}~\bibnamefont{Schutz}}, \bibnamefont{and}
  \bibinfo{author}{\bibfnamefont{H.}~\bibnamefont{Wilschut}},
  \bibinfo{journal}{Physics Letters B} \textbf{\bibinfo{volume}{538}},
  \bibinfo{pages}{27 } (\bibinfo{year}{2002}),
  \urlprefix\url{http://www.sciencedirect.com/science/article/pii/S0370269302019731}.

\bibitem[{\citenamefont{d'Enterria et~al.}(2001)\citenamefont{d'Enterria,
  Aphecetche, Chbihi, Delagrange, Diaz, van Goethem, Hoefman, Kugler, L\"ohner,
  Martinez et~al.}}]{thermal_photon_1}
\bibinfo{author}{\bibfnamefont{D.~G.} \bibnamefont{d'Enterria}},
  \bibinfo{author}{\bibfnamefont{L.}~\bibnamefont{Aphecetche}},
  \bibinfo{author}{\bibfnamefont{A.}~\bibnamefont{Chbihi}},
  \bibinfo{author}{\bibfnamefont{H.}~\bibnamefont{Delagrange}},
  \bibinfo{author}{\bibfnamefont{J.}~\bibnamefont{Diaz}},
  \bibinfo{author}{\bibfnamefont{M.~J.} \bibnamefont{van Goethem}},
  \bibinfo{author}{\bibfnamefont{M.}~\bibnamefont{Hoefman}},
  \bibinfo{author}{\bibfnamefont{A.}~\bibnamefont{Kugler}},
  \bibinfo{author}{\bibfnamefont{H.}~\bibnamefont{L\"ohner}},
  \bibinfo{author}{\bibfnamefont{G.}~\bibnamefont{Martinez}},
  \bibnamefont{et~al.}, \bibinfo{journal}{Phys. Rev. Lett.}
  \textbf{\bibinfo{volume}{87}}, \bibinfo{pages}{022701}
  (\bibinfo{year}{2001}),
  \urlprefix\url{https://link.aps.org/doi/10.1103/PhysRevLett.87.022701}.

\bibitem[{\citenamefont{Martinez et~al.}(1995)\citenamefont{Martinez, Marques,
  Schutz, Wolf, Diaz, Franke, Hlava, Holzmann, Lautridou, and
  Lefeve}}]{MARTINEZ199523}
\bibinfo{author}{\bibfnamefont{G.}~\bibnamefont{Martinez}},
  \bibinfo{author}{\bibfnamefont{F.}~\bibnamefont{Marques}},
  \bibinfo{author}{\bibfnamefont{Y.}~\bibnamefont{Schutz}},
  \bibinfo{author}{\bibfnamefont{G.}~\bibnamefont{Wolf}},
  \bibinfo{author}{\bibfnamefont{J.}~\bibnamefont{Diaz}},
  \bibinfo{author}{\bibfnamefont{M.}~\bibnamefont{Franke}},
  \bibinfo{author}{\bibfnamefont{S.}~\bibnamefont{Hlava}},
  \bibinfo{author}{\bibfnamefont{R.}~\bibnamefont{Holzmann}},
  \bibinfo{author}{\bibfnamefont{P.}~\bibnamefont{Lautridou}},
  \bibnamefont{and} \bibinfo{author}{\bibfnamefont{F.}~\bibnamefont{Lefeve}},
  \bibinfo{journal}{Physics Letters B} \textbf{\bibinfo{volume}{349}},
  \bibinfo{pages}{23 } (\bibinfo{year}{1995}),
  \urlprefix\url{http://www.sciencedirect.com/science/article/pii/037026939500236E}.

\bibitem[{\citenamefont{Zhang et~al.}(2017)\citenamefont{Zhang, Li, Wen, Liu,
  Su, and Zhang}}]{ZhangF}
\bibinfo{author}{\bibfnamefont{F.}~\bibnamefont{Zhang}},
  \bibinfo{author}{\bibfnamefont{C.}~\bibnamefont{Li}},
  \bibinfo{author}{\bibfnamefont{P.-W.} \bibnamefont{Wen}},
  \bibinfo{author}{\bibfnamefont{J.-W.} \bibnamefont{Liu}},
  \bibinfo{author}{\bibfnamefont{J.}~\bibnamefont{Su}}, \bibnamefont{and}
  \bibinfo{author}{\bibfnamefont{F.-S.} \bibnamefont{Zhang}},
  \bibinfo{journal}{Phys. Rev. C} \textbf{\bibinfo{volume}{100}},
  \bibinfo{pages}{024603} (\bibinfo{year}{2017}),
  \urlprefix\url{https://doi.org/10.1103/PhysRevC.100.024603}.

\bibitem[{\citenamefont{Feldmeier and Schnack}(2000)}]{FMD}
\bibinfo{author}{\bibfnamefont{H.}~\bibnamefont{Feldmeier}} \bibnamefont{and}
  \bibinfo{author}{\bibfnamefont{J.}~\bibnamefont{Schnack}},
  \bibinfo{journal}{Rev. Mod. Phys.} \textbf{\bibinfo{volume}{72}},
  \bibinfo{pages}{655} (\bibinfo{year}{2000}),
  \urlprefix\url{https://doi.org/10.1103/RevModPhys.72.655}.

\bibitem[{\citenamefont{Kerman and Koonin}(1976)}]{KERMAN1976332}
\bibinfo{author}{\bibfnamefont{A.}~\bibnamefont{Kerman}} \bibnamefont{and}
  \bibinfo{author}{\bibfnamefont{S.}~\bibnamefont{Koonin}},
  \bibinfo{journal}{Annals of Physics} \textbf{\bibinfo{volume}{100}},
  \bibinfo{pages}{332 } (\bibinfo{year}{1976}),
  \urlprefix\url{http://www.sciencedirect.com/science/article/pii/0003491676900658}.

\bibitem[{\citenamefont{Ono et~al.}(1992)\citenamefont{Ono, Horiuchi, Maruyama,
  and Ohnishi}}]{zero_energy}
\bibinfo{author}{\bibfnamefont{A.}~\bibnamefont{Ono}},
  \bibinfo{author}{\bibfnamefont{H.}~\bibnamefont{Horiuchi}},
  \bibinfo{author}{\bibfnamefont{T.}~\bibnamefont{Maruyama}}, \bibnamefont{and}
  \bibinfo{author}{\bibfnamefont{A.}~\bibnamefont{Ohnishi}},
  \bibinfo{journal}{Progress of Theoretical Physics}
  \textbf{\bibinfo{volume}{87}}, \bibinfo{pages}{1185} (\bibinfo{year}{1992}),
  \urlprefix\url{https://doi.org/10.1143/PTP.87.1185}.

\bibitem[{\citenamefont{Bertsch and Gupta}(1988)}]{BUU}
\bibinfo{author}{\bibfnamefont{G.}~\bibnamefont{Bertsch}} \bibnamefont{and}
  \bibinfo{author}{\bibfnamefont{S.~D.} \bibnamefont{Gupta}},
  \bibinfo{journal}{Physics Reports} \textbf{\bibinfo{volume}{160}},
  \bibinfo{pages}{189 } (\bibinfo{year}{1988}), ISSN \bibinfo{issn}{0370-1573},
  \urlprefix\url{http://www.sciencedirect.com/science/article/pii/0370157388901706}.

\bibitem[{\citenamefont{Xue et~al.}(2016)\citenamefont{Xue, Xu, Yong, and
  Ren}}]{ren_photon}
\bibinfo{author}{\bibfnamefont{H.}~\bibnamefont{Xue}},
  \bibinfo{author}{\bibfnamefont{C.}~\bibnamefont{Xu}},
  \bibinfo{author}{\bibfnamefont{G.-C.} \bibnamefont{Yong}}, \bibnamefont{and}
  \bibinfo{author}{\bibfnamefont{Z.}~\bibnamefont{Ren}},
  \bibinfo{journal}{Physics Letters B} \textbf{\bibinfo{volume}{755}},
  \bibinfo{pages}{486 } (\bibinfo{year}{2016}), ISSN \bibinfo{issn}{0370-2693},
  \urlprefix\url{http://www.sciencedirect.com/science/article/pii/S0370269316001404}.

\bibitem[{\citenamefont{Wiringa et~al.}(2014)\citenamefont{Wiringa, Schiavilla,
  Pieper, and Carlson}}]{momenta_distribution}
\bibinfo{author}{\bibfnamefont{R.~B.} \bibnamefont{Wiringa}},
  \bibinfo{author}{\bibfnamefont{R.}~\bibnamefont{Schiavilla}},
  \bibinfo{author}{\bibfnamefont{S.~C.} \bibnamefont{Pieper}},
  \bibnamefont{and} \bibinfo{author}{\bibfnamefont{J.}~\bibnamefont{Carlson}},
  \bibinfo{journal}{Phys. Rev. C} \textbf{\bibinfo{volume}{89}},
  \bibinfo{pages}{024305} (\bibinfo{year}{2014}),
  \urlprefix\url{https://link.aps.org/doi/10.1103/PhysRevC.89.024305}.

\bibitem[{\citenamefont{Hen et~al.}(2017)\citenamefont{Hen, Miller, Piasetzky,
  and Weinstein}}]{Hen}
\bibinfo{author}{\bibfnamefont{O.}~\bibnamefont{Hen}},
  \bibinfo{author}{\bibfnamefont{G.~A.} \bibnamefont{Miller}},
  \bibinfo{author}{\bibfnamefont{E.}~\bibnamefont{Piasetzky}},
  \bibnamefont{and} \bibinfo{author}{\bibfnamefont{L.~B.}
  \bibnamefont{Weinstein}}, \bibinfo{journal}{Rev. Mod. Phys.}
  \textbf{\bibinfo{volume}{89}}, \bibinfo{pages}{045002}
  (\bibinfo{year}{2017}),
  \urlprefix\url{https://link.aps.org/doi/10.1103/RevModPhys.89.045002}.

\bibitem[{\citenamefont{Hen et~al.}(2015)\citenamefont{Hen, Li, Guo, Weinstein,
  and Piasetzky}}]{hmt_Libaoan}
\bibinfo{author}{\bibfnamefont{O.}~\bibnamefont{Hen}},
  \bibinfo{author}{\bibfnamefont{B.-A.} \bibnamefont{Li}},
  \bibinfo{author}{\bibfnamefont{W.-J.} \bibnamefont{Guo}},
  \bibinfo{author}{\bibfnamefont{L.~B.} \bibnamefont{Weinstein}},
  \bibnamefont{and}
  \bibinfo{author}{\bibfnamefont{E.}~\bibnamefont{Piasetzky}},
  \bibinfo{journal}{Phys. Rev. C} \textbf{\bibinfo{volume}{91}},
  \bibinfo{pages}{025803} (\bibinfo{year}{2015}),
  \urlprefix\url{https://link.aps.org/doi/10.1103/PhysRevC.91.025803}.

\bibitem[{\citenamefont{Noll et~al.}(1984)\citenamefont{Noll, Grosse,
  Braun-Munzinger, Dabrowski, Heckwolf, Klepper, Michel, M\"uller, Stelzer,
  Brendel et~al.}}]{pion_velocity}
\bibinfo{author}{\bibfnamefont{H.}~\bibnamefont{Noll}},
  \bibinfo{author}{\bibfnamefont{E.}~\bibnamefont{Grosse}},
  \bibinfo{author}{\bibfnamefont{P.}~\bibnamefont{Braun-Munzinger}},
  \bibinfo{author}{\bibfnamefont{H.}~\bibnamefont{Dabrowski}},
  \bibinfo{author}{\bibfnamefont{H.}~\bibnamefont{Heckwolf}},
  \bibinfo{author}{\bibfnamefont{O.}~\bibnamefont{Klepper}},
  \bibinfo{author}{\bibfnamefont{C.}~\bibnamefont{Michel}},
  \bibinfo{author}{\bibfnamefont{W.~F.~J.} \bibnamefont{M\"uller}},
  \bibinfo{author}{\bibfnamefont{H.}~\bibnamefont{Stelzer}},
  \bibinfo{author}{\bibfnamefont{C.}~\bibnamefont{Brendel}},
  \bibnamefont{et~al.}, \bibinfo{journal}{Phys. Rev. Lett.}
  \textbf{\bibinfo{volume}{52}}, \bibinfo{pages}{1284} (\bibinfo{year}{1984}),
  \urlprefix\url{https://link.aps.org/doi/10.1103/PhysRevLett.52.1284}.

\end{thebibliography}

\end{CJK*}
\end{document}